\documentclass{JINST}
\usepackage{lineno}
\usepackage{amsmath}

\title{Measurements of proportional scintillation and electron multiplication in liquid xenon using thin wires}

\author{E.~Aprile$^a$, H.~Contreras$^a$, L.W.~Goetzke$^a$, A.J.~Melgarejo~Fernandez$^a$, M.~Messina$^a$, J.~Naganoma$^b$\thanks{Corresponding
author.}, G.~Plante$^a$, A.~Rizzo$^a$, P.~Shagin$^b$, and~R.~Wall$^b$\\
\llap{$^a$}Physics Department, Columbia University\\
  New York, NY 10027, USA\\
\llap{$^b$}Department of Physics and Astronomy, Rice University\\
  Houston, TX 77005, USA\\
  E-mail: \email{junji@rice.edu}}

\abstract{
Proportional scintillation in liquid xenon has a promising application in the field of direct dark matter detection, potentially allowing for simpler, more sensitive detectors. 
However, knowledge of the basic properties of the phenomenon as well as guidelines for its practical use are currently limited. 
We report here on measurements of proportional scintillation light emitted in liquid xenon around thin wires. 
The maximum proportional scintillation gain of $287\substack{+97\\-75}$ photons per drift electron was obtained using 10 $\mu$m diameter gold plated tungsten wire. 
The thresholds for electron multiplication and proportional scintillation are measured as $725\substack{+48\\-139}$ and $412\substack{+10\\-133}$ kV/cm, respectively. 
The threshold for proportional scintillation is in good agreement with a previously published result, while the electron multiplication threshold represents a novel measurement. 
A complete set of parameters for the practical use of the electron multiplication and proportional scintillation processes in liquid xenon was also obtained for the first time.
}


\begin{document}


\section{Introduction}
\label{sec:intro}

The search for dark matter continues to be an intense area of research within the worldwide scientific community.
A promising class of candidates for galactic dark matter are weakly-interacting massive particles (WIMPs), which calculations have shown would have approximately the correct relic abundance, assuming a GeV-scale mass and an interaction cross section on the weak scale.
Gas-liquid dual-phase time projection chambers (TPCs) using noble elements, xenon in particular, have emerged as leaders in this exciting field, and have set the strongest limits to-date on the WIMP-nucleon spin-independent elastic scattering cross section for a wide range of masses \cite{Aprile:2012nq, Akerib:2013tjd}.
Liquid xenon (LXe) is an attractive target for studying WIMP interactions with normal matter because it has no long-lived radioactive isotopes, and its physical and chemical properties make it ideal for the production and propagation of ionization electrons and scintillation photons \cite{Aprile:2010, Chepel:2013}.

Dual-phase TPCs rely on the simultaneous detection of direct scintillation photons produced in the liquid (the S1 signal) and electroluminescent photons produced in the gas proportional to the number of ionization electrons extracted from the liquid (the proportional scintillation or S2 signal).
From the time difference between the S1 and S2 signals and the localized position of the S2 signal, one can determine the location of the initial interaction, allowing for a fiducialization of the detector volume to take advantage of LXe's excellent stopping power for external background particles.
Furthermore, the ratio of the S1 and S2 signals is substantially different for nuclear and electronic recoils, allowing for discrimination between WIMP-like interactions and those originating from electromagnetic interactions. 
These two background rejection methods have been successfully used to make xenon TPCs among the best low background detectors for the measurement of rare nuclear recoils.

An increase in sensitivity beyond that achieved by current xenon experiments requires detectors with a much larger active target mass, at  least an order of magnitude larger than the current, one hundred kilogram scale, state of the art.
Within the XENON dark matter project, the first dual-phase TPC with a ton scale fiducial Xe target is being realized. 
A simple scale-up of a dual-phase TPC presents practical challenges and difficulties, most notably associated with high voltage handling and precise control of the liquid-gas interface level, as the total drift length and cross-sectional area of the TPC, respectively, increase.
These challenges would be much reduced in a single-phase LXe TPC, as such a detector would allow for the implementation of two or more drift gaps with a reduced high voltage requirement,  and would not require precise liquid level control~\cite{Giboni:2011}. 
Without total internal reflections suffered by the primary light at the liquid-gas interface, a single-phase TPC will also have a higher intrinsic light collection efficiency for the S1 signal.  
The main challenge would then become producing a S2 signal with sufficient gain.

The feasibility of proportional scintillation in LXe was first demonstrated by Lansiart et al. in 1976 using thin wires \cite{bib:Lansiart1}, and confirmed a few years later by Masuda et al. \cite{bib:Masuda1}.
Within the XENON dark matter project, we therefore embarked on a study to reproduce and further understand the phenomenon in view of its potential application for a next generation, large scale LXe dark matter detector. 
This paper presents our first results with proportional scintillation in LXe. 
The paper also addresses questions relevant for the practical application of proportional scintillation in LXe for a  dark matter detector, namely the number of proportional scintillation photons produced per electron, the electric field thresholds of proportional scintillation and electron multiplication~\cite{bib:Derenzo1, bib:Policarpo1}, and the resolution on electron counting.

It is also important to point out that other schemes beyond thin wires are possible and have recently been tested for the production of proportional scintillation in LXe in \cite{bib:Arazi}.
Here, the authors used a gas electron multiplier~(GEM)~ instead of the thin wire used in our work.
GEMs offer the potential advantage, among others, of a larger gain, and thus could offer greater sensitivity to WIMP searches \cite{bib:Buzulutskov}.

The paper is organized as follows: we describe the setup of the experimental apparatus in Section~\ref{sec:exp}, with details on the data acquisition, calibration, and experimental procedures given separately.  
Details on the analysis technique, simulations, and event selections are given in Section~\ref{sec:ana}.  
Finally, the results of the proportional scintillation measurements and the associated systematic uncertainties, are given in Section~\ref{sec:res}, followed by a discussion of these results and future prospects in Section~\ref{sec:dis}.
\section{Experimental setup}
\label{sec:exp}

The detector developed for this study is a single-phase TPC, built with four grid electrodes supported by a polytetrafluoroethylene~(PTFE) structure, and viewed by two photo-multiplier tubes (PMTs), one located at the top of the TPC and one at the bottom, all immersed in LXe.
The TPC was irradiated with $\alpha$-particles emitted by a $^{210}$Po source deposited on the cathode grid, producing direct scintillation light and ionization electrons.
The electric field produced between the cathode and the gate grid drifts the electrons towards the anode wire, where a strong field accelerates them, leading to the S2 signal in LXe.
A charge-sensitive preamplifier was coupled to the anode frame to directly measure the number of electrons produced in the interaction.
The TPC was cooled by a cold finger immersed in liquid nitrogen (LN$_2$).

\begin{figure}[htbp]
  \begin{center}
    \includegraphics[width=.45\textwidth]{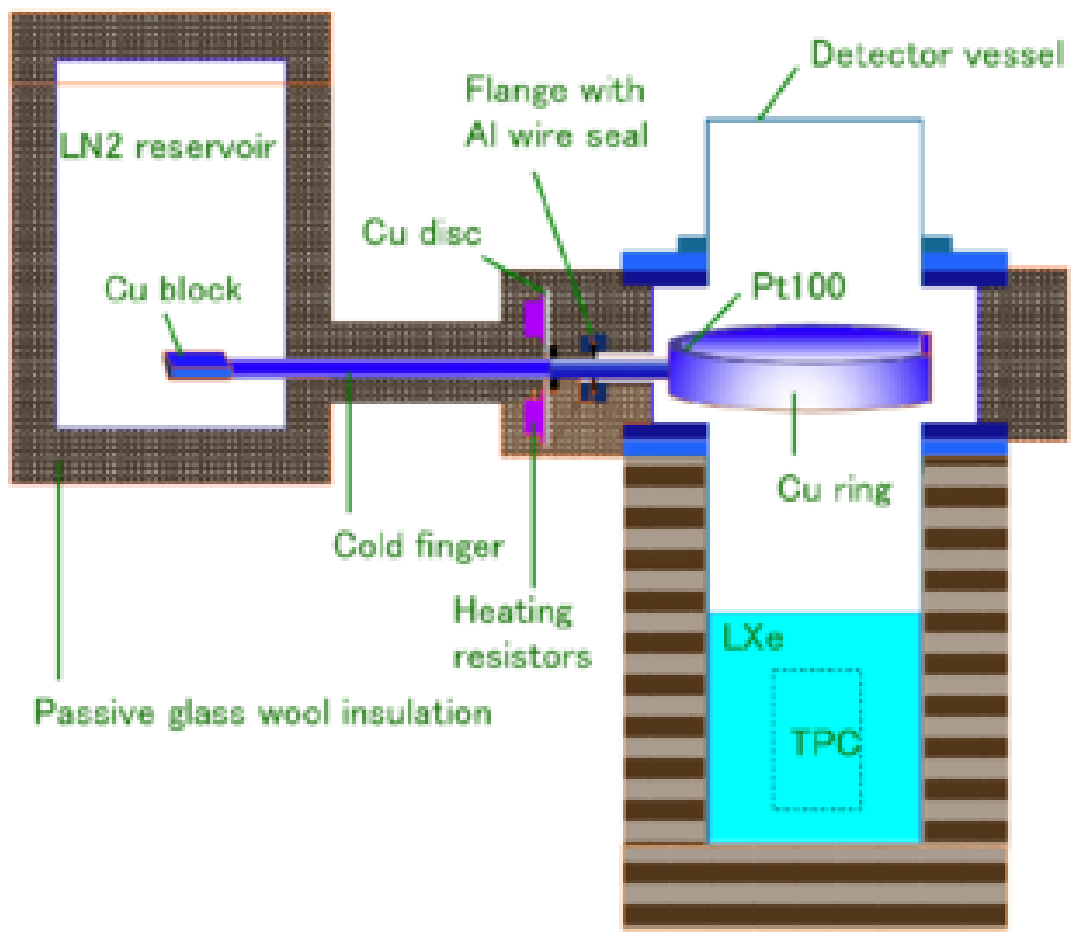} 
    \includegraphics[width=.45\textwidth]{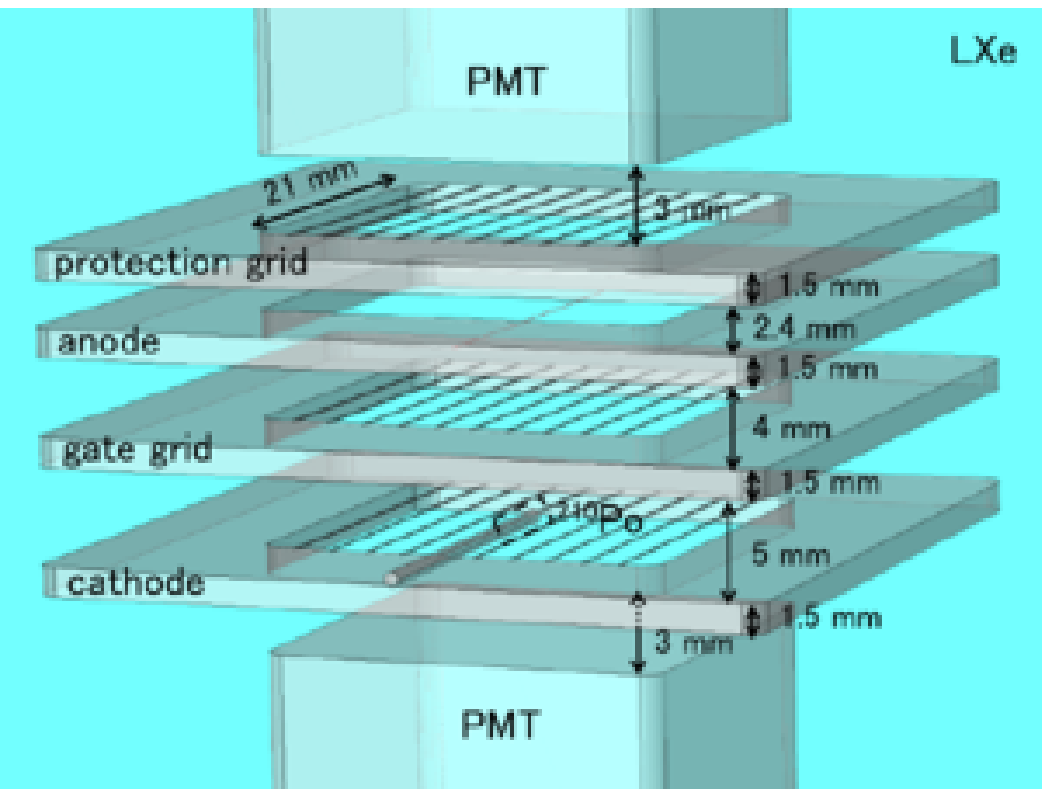} 
    \caption{Left: a schematic drawing of the cooling system.
      Right: a schematic drawing of the single-phase LXe TPC. 
      A thin wire is placed on the anode frame.}
    \label{fig:setup}
  \end{center}
\end{figure}

The xenon was continuously purified using a high temperature getter, throughout the duration of each data taking period.
A schematic drawing of the cooling system, and TPC used in the experiment are shown in figure~\ref{fig:setup}.
The setup was located at Columbia University's Nevis Laboratories.
The rest of this section is devoted to giving additional details about the experimental setup.

\subsection{Detector, purification, and cooling}
\label{sec:apparatus}

The active LXe volume of 21 x 21 x 23 mm$^{3}$ was viewed by two high quantum efficiency (QE) Hamamatsu R8520-406 SEL PMTs~\cite{hamamatsu} with square bialkali photocathodes measuring 21x21 mm$^{2}$ in cross section, located at the top and bottom of the TPC.
These are the same type of PMTs used in the XENON100 dark matter experiment \cite{Aprile:2011dd}.
The photocathodes were designed for operation down to -110 $\ensuremath{^\circ}$C, well below the temperature where xenon is in the liquid phase (-95 $\ensuremath{^\circ}$C).
Measurements reported in Ref. \cite{Aprile:2012dy} show the PMTs have an average LXe temperature QE of 33\% at 178 nm, the wavelength at which the xenon scintillates.
A CAEN SY4527 high voltage system~\cite{CAENHV} was used to supply high voltage to both PMTs.
By design, the PMTs may be operated up to 1 kV, but in most operating conditions a voltage of +750 V or less was applied.
A positive bias voltage was used to keep the metal body and photocathode grounded.
The calibration procedure for the system is described in Section \ref{sec:calib}.

Four grid electrodes (the protection grid, anode, gate grid, and cathode) were used to provide an electric field in the drift region, and a much stronger electric field near the anode wire to produce the proportional scintillation signal. 
The protection grid, gate, and cathode were each made of nine stainless steel wires with a diameter of 100~$\mu$m.
The wires were soldered to the surface of a 38 x 38 x 1.5 mm stainless steel frame with a 2.1 mm pitch.
A stainless steel needle, 550 $\mu$m in diameter, was soldered to the cathode frame, just above the center of the cathode grid.
A $^{210}$Po source was deposited in a 3.5 mm strip around the eye of the needle to serve as a source of $\alpha$-particles.
The anode electrode consisted of one gold plated tungsten wire 10 $\mu$m in diameter welded at the center of the stainless steel frame.
Measurements were also made with a 5 $\mu$m anode wire.
High voltage was separately applied to the anode, gate, and cathode using the same CAEN high voltage power supply used for the PMTs.
The protection grid electrode was grounded to close the electric field inside the active volume.

An Amptek charge-sensitive preamplifier~\cite{AMPTEK} with a shaping amplifier (3.5 $\mu$s shaping time) was connected to the anode, to allow for the measurement of the charge signal. 
The PMT bases and connections were shielded with copper foil to reduce pick up noise from the PMT signal, and low pass filters were used to reduce noise from the power supply. 
Finally, a high pass filter was used to isolate the preamplifier input from the high voltage supply.

A PTFE support structure was used to hold the PMTs, reflect the scintillation light, and provide the proper spacing between grid electrodes. 
Using these spacers, the distance between the cathode and gate wires was set to 6.5 mm, while the distance between the gate and the anode wires was 4.0 mm, and the distance between the anode and the protection grid wires was 3.9 mm.
The PTFE structure was suspended  using a stainless steel rod, mounted on the top flange of the vessel containing the TPC. 
The custom-designed, insulated vessel is the same as used for the measurements described in \cite{bib:Guillaume1, bib:Kyungun1}.


Although pulse tube refrigerators have proven to be the easiest and most reliable way of cooling a LXe detector, a cryo-cooler is not always available. 
Liquefying the xenon gas with a LN$_2$ coil was historically used very frequently. 
In such set-ups, a pressure controller switches the LN$_2$ flow and keeps the pressure and thus the temperature of the liquid within range. 
However, for dual-phase systems, the gain of the proportional scintillation depends on the gas pressure, i.e. the signal has to be corrected with constantly changing factors depending on the actual pressure value.
The present cooling system uses LN$_2$, in thermal contact with the active volume via a solid oxygen-free high thermal conductivity copper cold finger, as a cold source.
A schematic of the cooling system is shown in figure \ref{fig:setup} (left). 
The LN$_2$ is stored in a large stainless steel vessel with passive glass wool insulation.                                      
The vessel is equipped with two level sensors to control the filling of additional LN$_2$, when required.
A solid copper block with large surface area is immersed in the LN$_2$ and is placed in thermal contact with a solid 1'' diameter copper rod. 
The copper rod exits the reservoir on the side and enters a circular section of the detector vessel above the active TPC region.
The cold finger is face sealed to a custom stainless steel flange. 
Since an indium sealing would limit the bake out temperature, we chose an aluminum wire seal consisting of a ring of 1/16'' pure, soft aluminum. 
A circular shaped copper bar, 1.5'' high and 0.25'' thick connects to the end of the cold finger and is concealed in a recessed area close to the walls. 
This keeps ample free space in the center for cabling. 
The xenon gas is condensed to the liquid phase via the thermal gradient across the cold finger.
When the copper ring reaches the appropriate temperature, the gaseous xenon liquefies and drips into the TPC volume. 
To monitor the temperature on the cold finger, a Pt100 resistor is mounted on the copper bar close to the junction with the cold finger. 
Eight power resistors mounted on a 4'' diameter copper disc installed outside of the detector vessel and in thermal contact with the cold finger provide heat, as necessary, to keep the temperature on the cold finger above that at which the xenon will freeze.
A proportional-integral-derivative controller~\cite{LAKESHORE} can counteract excessive cooling power by adjusting the power dissipation in the resistors depending on the temperature of the copper bar in the vessel. 
All the cooling structure, including the LN$_2$ reservoir, are thermally insulated by about 3'' of glass wool.
The cooling power of this setup was measured to be approximately 70~W at the typical operating temperature of -95~$\ensuremath{^\circ}$C.


The gas purification system used for the experiment is the same as described in~\cite{bib:Guillaume1}.
Xenon was continuously purified by a hot getter \cite{SAES} to remove electronegative impurities in the liquid.
These impurities capture the ionization electrons, and thus degrade the size of the S2 signal.
The purification cycle began when LXe was vaporized and passed through the hot getter.
Purified gas returning from the getter was re-condensed on the copper bar, and collected by an aluminium funnel.
A plastic tube connected to the funnel guides the purified LXe directly into the TPC, allowing for a faster introduction of the pure xenon into the sensitive region of the detector.

\subsection{Data acquisition}
\label{sec:daq}
\begin{figure}[htbp]
\begin{center}
\includegraphics[width=.6\textwidth]{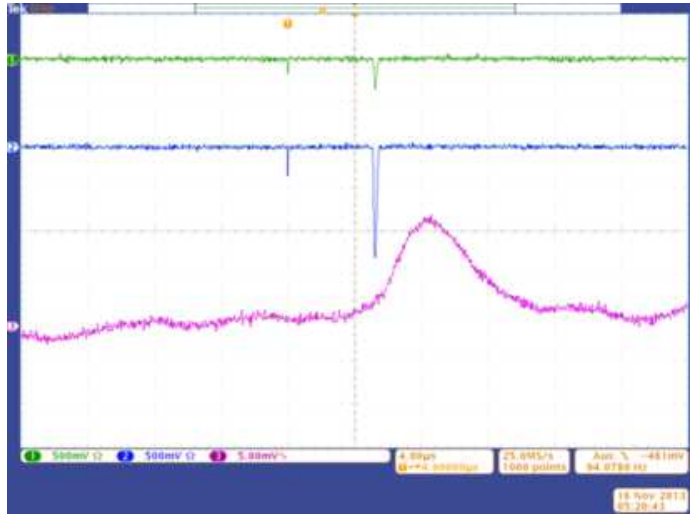}
\caption{
  Oscilloscope waveforms from a typical $\alpha$-particle interaction in LXe.
  Green, blue, and purple waveform traces represent the top PMT, bottom PMT, and charge-sensitive preamplifier signals, respectively. 
  Time difference between the S1 and S2 signals is compatible with the drift time of the electrons from the cathode to the anode.
  The voltage difference between the gate electrode and anode was 6.5~kV, and that between the cathode and the gate was 0.5~kV. 
}
\label{fig:waveform}
\end{center}
\end{figure}

The TPC was irradiated with $^{210}$Po.
The range of 5.4~MeV $\alpha$-particles in LXe is less than 50~$\mu$m, hence interactions are very localized around the $^{210}$Po source.
The S1 signal was detected by the top and bottom PMTs and provides a trigger signal for the data acquisition system~(DAQ). 
Assuming that the average energy to produce an electron-ion pair ($W$ value) is $15.6$~eV~\cite{bib:Takahashi} and the fraction of electrons escaping recombination of $4~\%$~\cite{Aprile:1991}, it was inferred that about 14,000 electrons were liberated from the $\alpha$-particle interaction.
Under a $\sim 1$~kV/cm drift field provided by the cathode, these electrons drift towards the proportional region with an average drift velocity of 1.75~mm/$\mu$s~\cite{bib:Miller}. 
After passing the gate electrode, electrons were accelerated in a region with high electric field provided by the anode wire.  
The S2 signal produced in this region was measured by the PMTs, and the charge signal was measured by the preamplifier.
An example of the S1 and S2 waveforms from the PMTs and the charge signal from the preamplifier, observed with an oscilloscope, is shown in figure~\ref{fig:waveform}.

The signals from the top and bottom PMTs were fed to a 50~$\Omega$ impedance signal divider with two outputs per channel. 
The first output was digitized by a CAEN flash ADC \cite{CAEN} with 2.25~V$_{pp}$ 14-bit and 100 Msamples/s.
Recorded waveforms were stored for offline analysis.
The charge signal from the preamplifier was digitized by the same flash ADC.
The second output was connected to the input of a leading edge discriminator~\cite{LECROY} with a $-30$~mV threshold, and used to select events of particular interest.
During normal data taking, the signal from the top PMT was used as the trigger for the DAQ.

\subsection{Calibration}
\label{sec:calib}

Gain calibration of the PMTs was performed regularly using a light-emitting diode (LED). 
The gain of each PMT was measured by fitting the single photo-electron~(PE) spectrum produced by illuminating the PMT photocathode with low-intensity pulses from the LED.
The LED was controlled by a pulse generator~\cite{BNC}, which also provides a trigger for the DAQ system.
The gains of the two PMTs varied between $4.2\times10^5$ to $4.2\times10^6$, depending on the high voltage setting.

\begin{figure}[htpb]
\begin{center}
\includegraphics[width=.5\textwidth]{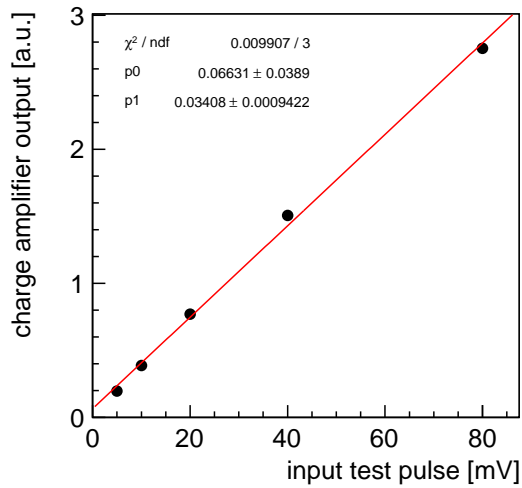}
\caption{Charge-sensitive preamplifier response as a function of input test pulse height.}
\label{fig:A225_calib}
\end{center}
\end{figure}

Calibration of the charge-sensitive preamplifier was performed using a pulse generator.
For charge calibration of the readout electronics, a precision generator~\cite{SRS} with a sharp rise time was used to inject a known amount of charge into the preamplifier through a calibrated 1~pF capacitor. 
The equivalent noise charge for the preamplifier in most conditions was approximately 1,450~electrons.
The linearity of the preamplifier output is shown in figure \ref{fig:A225_calib}.

\subsection{Experimental procedure}
Once the TPC was installed and a stable vacuum of at least $1.0\times10^{-5}$~mbar was achieved, the vessel was filled with gaseous xenon, which was subsequently condensed by the LN$_2$-based cooling system.
A total of 1.3~kg of xenon was used to fill the vessel and completely cover the top PMT window with LXe.
After filling, the xenon was purified at a recirculation speed of 2.8 standard liters per minute for at least 48 hours.
At this point, it was observed that the average S2 signal amplitude did not grow appreciably with increased recirculation time, and the data taking was started.
Recirculation was run continuously during each data taking period.

In each experiment, the primary measurement was to track the evolution of the S2 signal as a function of the field strength in the high electric field region.
Along these lines, the voltage difference between the anode and gate ($V_{A}$) was varied from 0.40~V (0.40~V) to 6.75~(3.00)~kV for studies with an anode wire diameter of 10~(5)~$\mu$m.
Continuous photon emission was observed at higher electric fields.
While the gate grid was held at ground potential for $V_{A} \leq 5.5$~kV, negative voltages were applied to the gate grid to reach higher values of $V_{A}$.
The voltage difference between the gate and cathode electrodes ($V_{C}$) was kept constant to produce a $\sim 1$~kV/cm electric field in the drift region for most values of $V_{A}$.
At smaller values of $V_{A}$, a weaker drift field was applied to prevent the capture of drift electrons by the gate wires.
To account for this effect in the data analysis, a scale factor for the number of electrons produced by the $\alpha$-interaction was derived using data collected $V_{A}$ = 3.5 kV for a number of different drift fields, and applied to those data samples at lower drift fields.

The size of the measured S2 signals varied over five orders of magnitude, well above the dynamic range of the flash ADC.
To prevent saturation of the ADC at high values of $V_{A}$, lower PMT voltages and an attenuator were used, as necessary, to reduce the size of the S2 signal amplitude.
The S2 signal size at the reference value of $V_{A} = 3.5$~kV was used to monitor the stability of the detector's response for the 10~$\mu$m wire study.
A corresponding reference value of $V_{A} = 3.0$~kV was used for the 5~$\mu$m wire study.
Regular PMT gain and charge-sensitive preamplifier calibrations were done before and after the measurements.
\section{Analysis}
\label{sec:ana}

Recorded digitized waveforms from the top and bottom PMTs, together with the charge signal, were analyzed to study the properties of proportional scintillation in LXe.
Event selection criteria were applied to the data in order to reject background events coming from cosmic ray interactions and other environmental sources, and to select good events coming from $\alpha$-particle interactions in the LXe. 
In order to fully characterize the proportional scintillation process, a number of quantities must also be derived using theoretical calculations or simulations.  
For example, to determine the number of proportional scintillation photons produced in the interaction, it is necessary to calculate the S2 light collection efficiency for the top and bottom PMTs. 
The precise electric field map and the experimental light collection efficiencies were studied in detail using COMSOL and Geant4.
In this section, details of the event selection and results of the simulations are presented.

\subsection{Simulation}
\label{sec:sim}

\begin{figure}[htbp]
 \begin{center}
  \includegraphics[width=.4\textwidth]{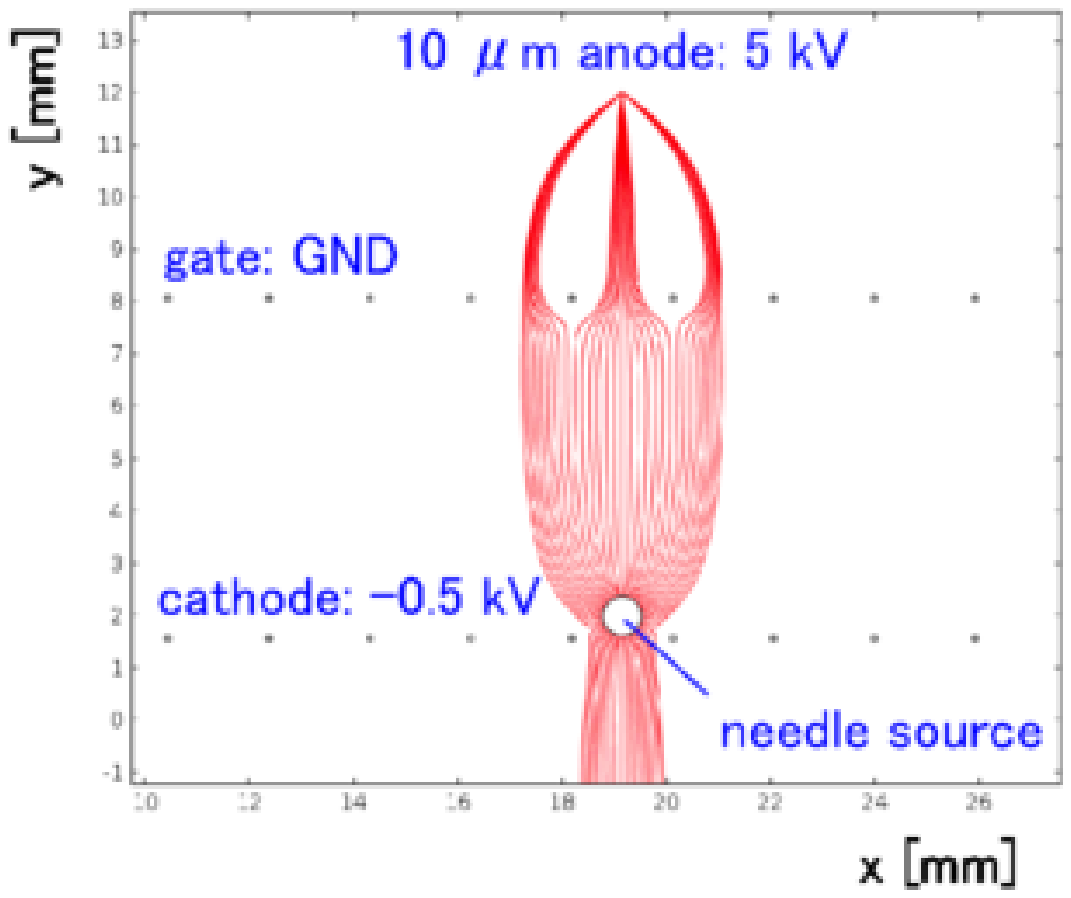} 
  \includegraphics[width=.5\textwidth]{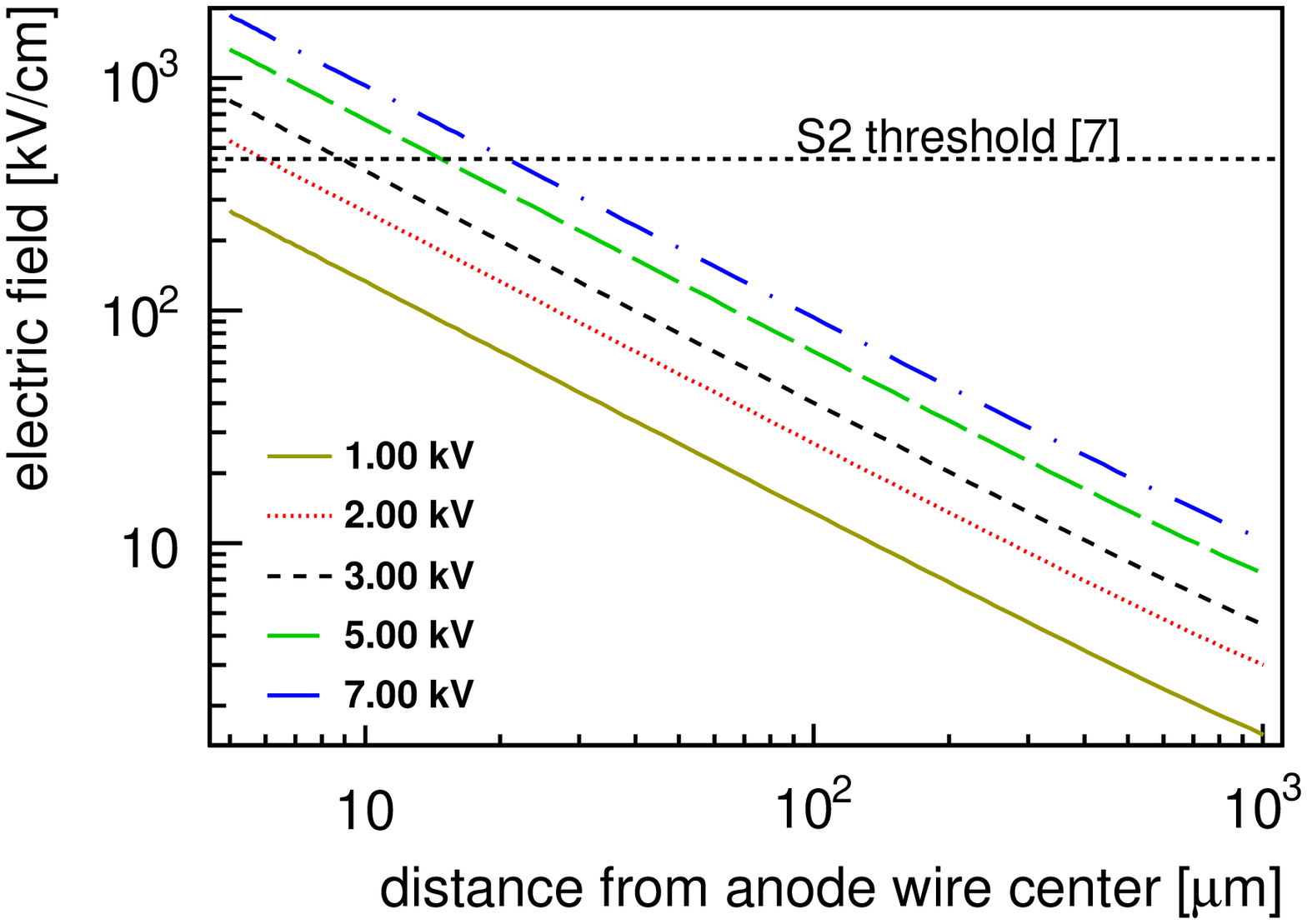} 
  \caption{Left: The result of the electric field simulation for the drift region. 
    Electron drift lines from the needle source are shown in red. 
    Three main drift paths towards the anode wire are clearly visible. 
    No streamlines are present which end at the gate electrode wires.
    Right: Value of the electrical field as a function of the distance from the anode wire center (10 $\mu$m diameter wire). 
    Solid, dotted, dashed, long-dashed and long-dashed-dotted lines correspond to $V_{A}$  = 1, 2, 3, 5, and 7 kV, respectively. 
    A horizontal dashed line represents the S2 threshold from \cite{bib:Masuda1}.}
  \label{fig:COMSOL}
 \end{center}
\end{figure}

The COMSOL multi-physics simulation package~\cite{comsol} was used to model the electric field inside the detector for a variety of voltage configurations.
The results of the electric field simulation in the drift region and the area close to the anode wire are shown in figure~\ref{fig:COMSOL}.
Three possible drift paths are clearly visible, two of which have equal drift length.
A small misalignment of the needle source with respect to the other electrodes breaks the symmetry of the two longer drift paths shown in figure~\ref{fig:COMSOL} (left), and result in three distinct drift paths.
The drift time resolution (10 ns) was sufficient to separately resolve these three paths, as the drift length difference between the central and outer paths was approximately 800 ns.
Only the central path was chosen to minimize the effect of electrode misalignment.
The electric field below the anode wire, shown in figure~\ref{fig:COMSOL} (right), was used to fit the data.

\begin{figure}[htbp]
 \begin{center}
  \includegraphics[width=.5\textwidth]{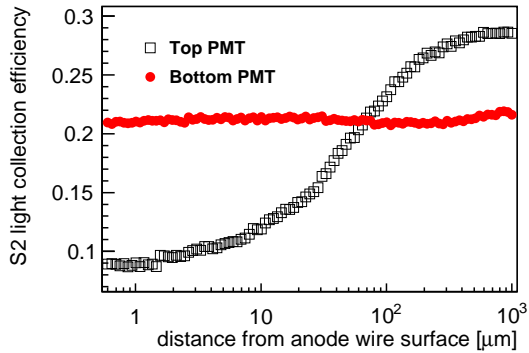} 
  \caption{LCE of the S2 signal for the top (empty squares) and bottom (full circles) PMTs as a function of the distance from the anode wire surface (10 $\mu$m diameter wire). 
    While LCE for the bottom PMT is almost independent of the distance to the anode surface, the one for the top PMT decreases near the anode due to shadowing of the scintillation light by the anode wire surface itself.}
  \label{fig:LCE}
 \end{center}
\end{figure}

To obtain the absolute number of emitted S2 photons, a dedicated simulation was used to estimate the light collection efficiency (LCE) of the two PMTs.
The Geant4 toolkit~\cite{geant4} was used to model the LCE for both top and bottom PMTs in a small region just below the anode wire.
The procedure used to model the system is similar to that described in ~\cite{bib:Guillaume1} for a similar detector. 
The simulated LCE for the top and bottom PMTs is shown in figure~\ref{fig:LCE}.
No significant difference between 10~$\mu$m and 5~$\mu$m anode wires was found.
The LCE for the bottom PMT is approximately 21\% and mostly independent of the distance from the anode wire.
Shadowing of the scintillation light by the anode wire itself leads to the observed drop in LCE for the top PMT in the region closest to the anode wire.
For these reasons, the S2 signal measured by the bottom PMT was used for the analysis. 

\subsection{Event selections}
\begin{figure}[htbp]
 \begin{center}
  \includegraphics[width=.5\textwidth]{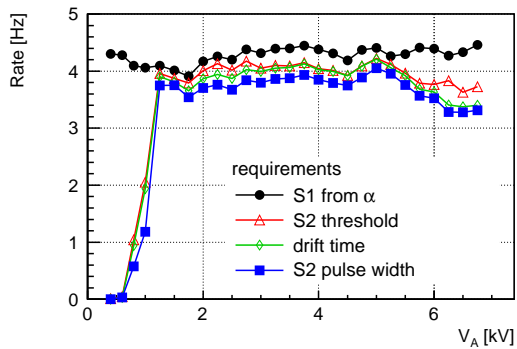} 
  \caption{Dependence of the counting rates on the value of the anode voltage $V_{A}$ for different event selection criteria for data collected with the 10 $\mu$m diameter anode wire. 
    No significant bias on the event selection criteria can be seen for $V_{A}$ $>$ 1.25 kV. 
    The efficiency drops at the lower voltage differences due to a decreased S2 signal size.}
  \label{fig:evtrate}
 \end{center}
\end{figure}

Several selections were applied to the PMT signals to reject different sources of background (cosmic rays, environmental, etc.) and isolate events originating from $\alpha$-particle interactions.
First, the size of the S1 signal must be compatible with the expected signal size from an $\alpha$-particle interaction in the LXe. 
Further, we select events with a light ratio between the top and bottom PMTs above $0.5$, in order to reject events on the bottom or side of needle source.
Second, the pulse height of the S2 signal must be more than 2~mV to reject electronic noise and/or thermal electron emission from the PMT photocathode.
Third, the time difference between the S1 and the S2 signals must be consistent with the drift distance between the anode and the cathode.
Last, the S2 signal time width must be consistent with the lone central drift path (see figure \ref{fig:COMSOL}) to reject events in which other drift paths are used.
The S2 width at 10~\% of pulse height must be less than about 450~ns depending on the $V_A$.
The dependence of the counting rates on $V_{A}$ for each individual event selection criteria is shown in figure~\ref{fig:evtrate}.
While no significant change in the counting rates can be seen for $V_{A}$  $>$~1.25~kV, the event selection rate drops at lower voltages due to a smaller average S2 signal size. 

\begin{figure}[htbp]
 \begin{center}
  \includegraphics[width=.48\textwidth]{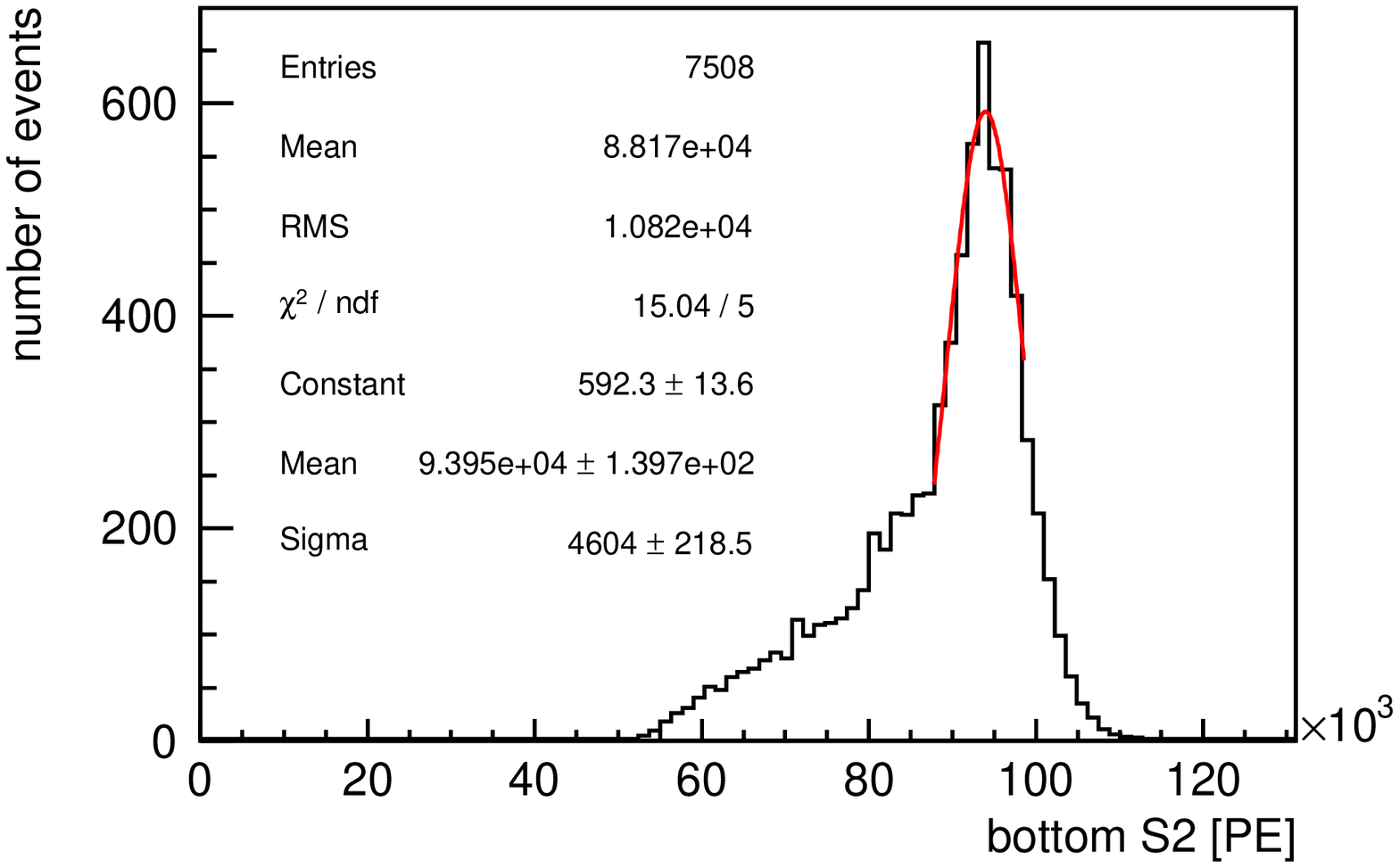} 
  \includegraphics[width=.48\textwidth]{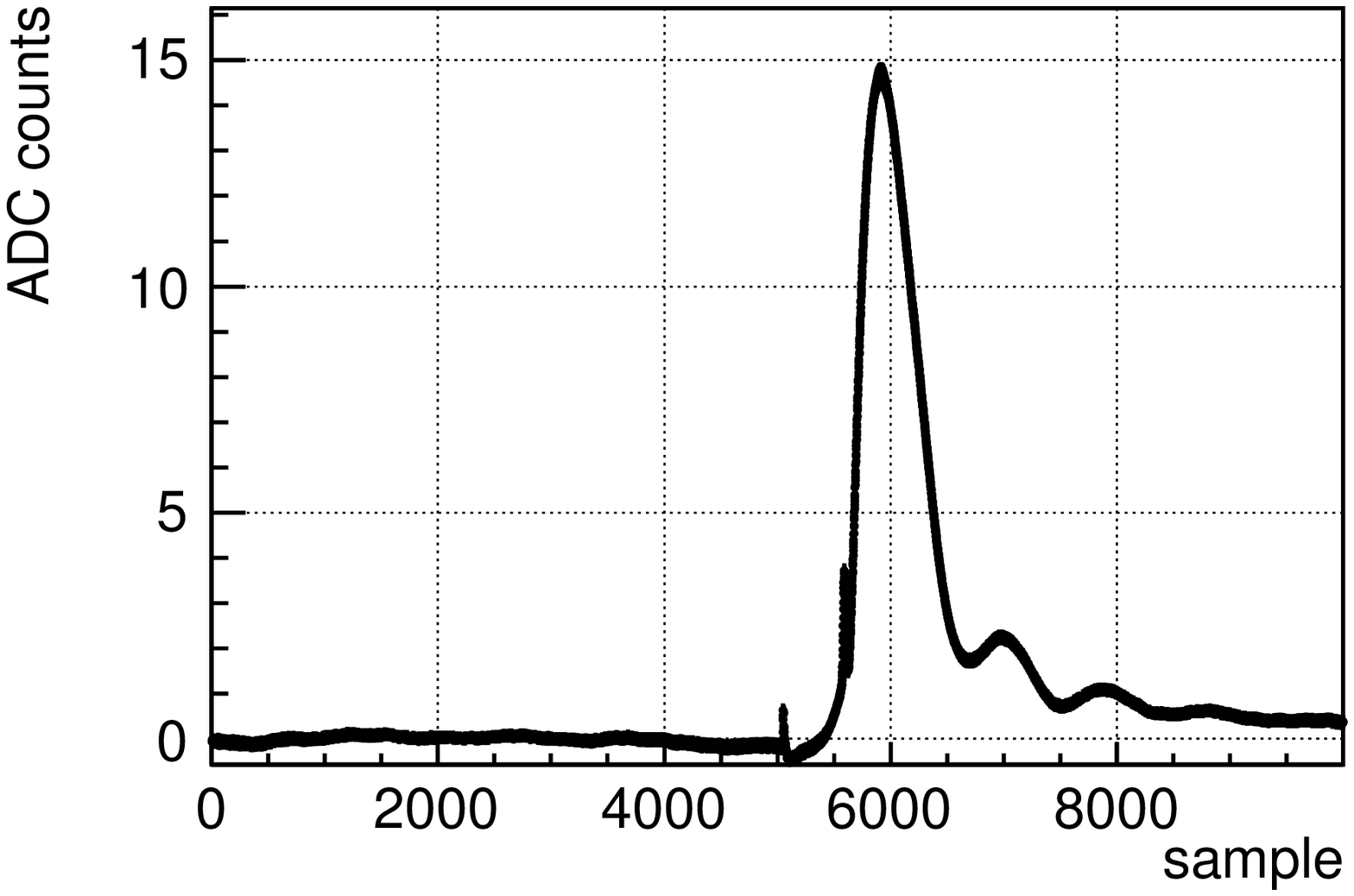} 
  \caption{A typical result obtained using the 10 $\mu$m diameter anode wire with $V_{A} = 5.00$~kV.
    Left: S2 signal distribution after the event selection criteria applied. 
    The peak of this distribution is fit with a Gaussian to define its mean and width.
    Right: Averaged charge signal after the event selection. 
    The small peaks seen at 5000 and 5600 samples are due to pick up noise induced by the S1 and S2 signals of the PMTs.}
  \label{fig:dist5kV}
 \end{center}
\end{figure}

The bottom PMT S2 distribution and the averaged charge signal after application of the event selection criteria are shown in figure~\ref{fig:dist5kV}.
Preamplifier waveforms were averaged over many events to improve the signal to noise ratio.

Waveforms were processed for offline analysis by an algorithm based on the one used by XENON100 \cite{Aprile:2011dd}.
Further modifications were needed for $V_{A} \leq 1.00$~kV to compensate for the inefficiency of the event selection in this region: no minimum threshold on the size of the S2 signal was applied, and the largest peak (regardless of size) around the expected drift time was taken to constitute the S2 signal.
Some contributions from noise are expected with this modification, but minimal background is expected in the normal selection (i.e. for $V_{A} \geq 1.25$~kV).

\section{Results}
\label{sec:res}
In this section, the main analysis results are presented.
First, the main sources of systematic uncertainty are discussed.
Next, the total S2 gain for this setup is calculated using the observed S2 signal.
A theoretical model is then used to establish the relationship between the number of drift electrons, the number of proportional scintillation photons, the local electric field along the electron drift path, and the parameters describing the proportional scintillation process.
By fitting this model to the data collected with the 10 and 5 $\mu$m wires, we extract the thresholds for proportional scintillation and charge multiplication.
Finally, results are presented related to the S2 signal width and resolution, with a discussion on the proportionality of the observed S2 signal.

\subsection{Systematic uncertainties}
Two main classes of systematic uncertainty are relevant for this analysis.
The first class consists of uncertainties which are related to the size of the measured S2 and charge signals, and thus affect the fit used to extract the thresholds. 
Uncertainties in this class are evaluated both for the PMT signal and the preamplifier.
With one exception, these uncertainties are evaluated using measurements performed at each value of $V_{A}$.
The second class of uncertainty is related to the conversion factor (described in detail later in this section) which is used to convert the observed S2 signal in photo-electrons to the gain in photons per drift electron ($ph/e^-$).

First, we discuss those uncertainties related to the fit, which are evaluated using measurements performed at different voltage conditions. 
Examples of this class of uncertainty related to the PMT signals are: the linearity of the PMT response, the PMT gain calculation, the scale factor at low $V_{A}$ due to different drift field conditions, and the non-Gaussian shape of the S2 distribution.
Examples of this class of uncertainty related to the preamplifier are: the scale factor used at low $V_{A}$, the linearity of the calibration, and the baseline evaluation and averaging.

The linearity of the PMT response is assessed using data taken with an LED of variable intensity for a fixed PMT voltage.
The linearity of the preamplifier is calculated by comparing the observed charge signal for different known input test pulses.
The uncertainty on the PMT gain calibration is assessed by comparing different binning structures and fitting methods to perform the calibration.
For each value of $V_{A}$, the mean value of the S2 signal size distribution is calculated using a Gaussian fit.
Particularly at lower values of $V_{A}$, this distribution acquires large non-Gaussian tails, as indicated in figure \ref{fig:dist5kV} (left) for data taken at $V_{A}$ = 5.0 kV.
To assess the effect of these tails on the calculation of the peak position of the S2 signal, the mean of the distribution is compared to the fitted peak.
The uncertainty on the low-$V_{A}$ scale factor is determined by the peak-finding uncertainty used in the derivation of the scale factor itself.

The final uncertainty affecting the fit is related to the overall stability of the measurement over the course of the data-taking program.
To assess this uncertainty, we consider a series of data samples taken with the same voltage configuration ($V_{A}$~=~3.5~(3.0)~kV for the 10~(5)~$\mu$m wire) over the course of 5 days.
For the PMT signal, the average size of the S2 signal is calculated for each run, and the spread of this distribution is taken as a measure of the overall stability.
The same data are used to assess the stability of the preamplifier by comparing the size of the observed charge signal.
The value of this uncertainty is assumed to hold for all voltage configurations.

\begin{figure}[htbp]
 \begin{center}
  \includegraphics[width=.48\textwidth]{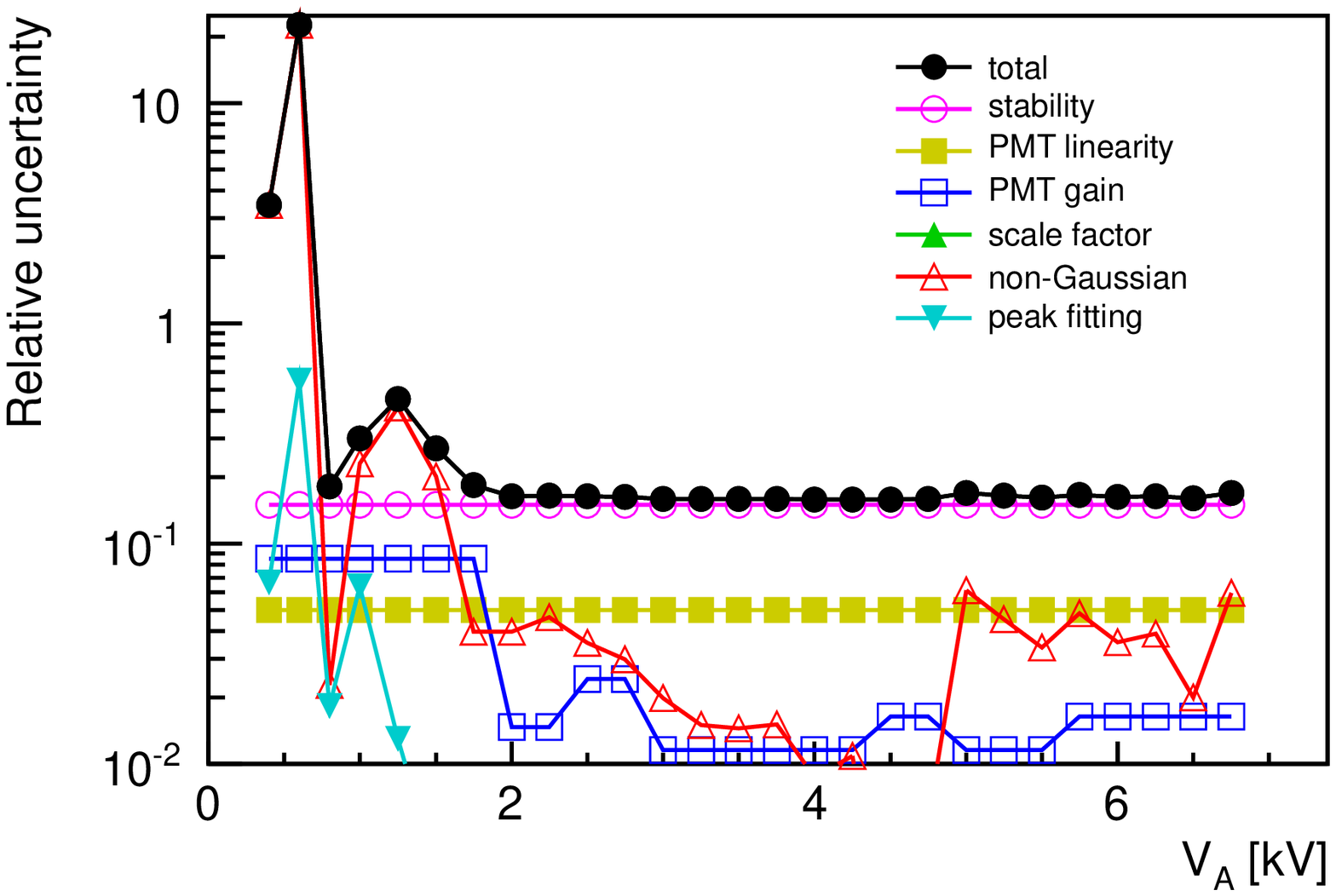} 
  \includegraphics[width=.48\textwidth]{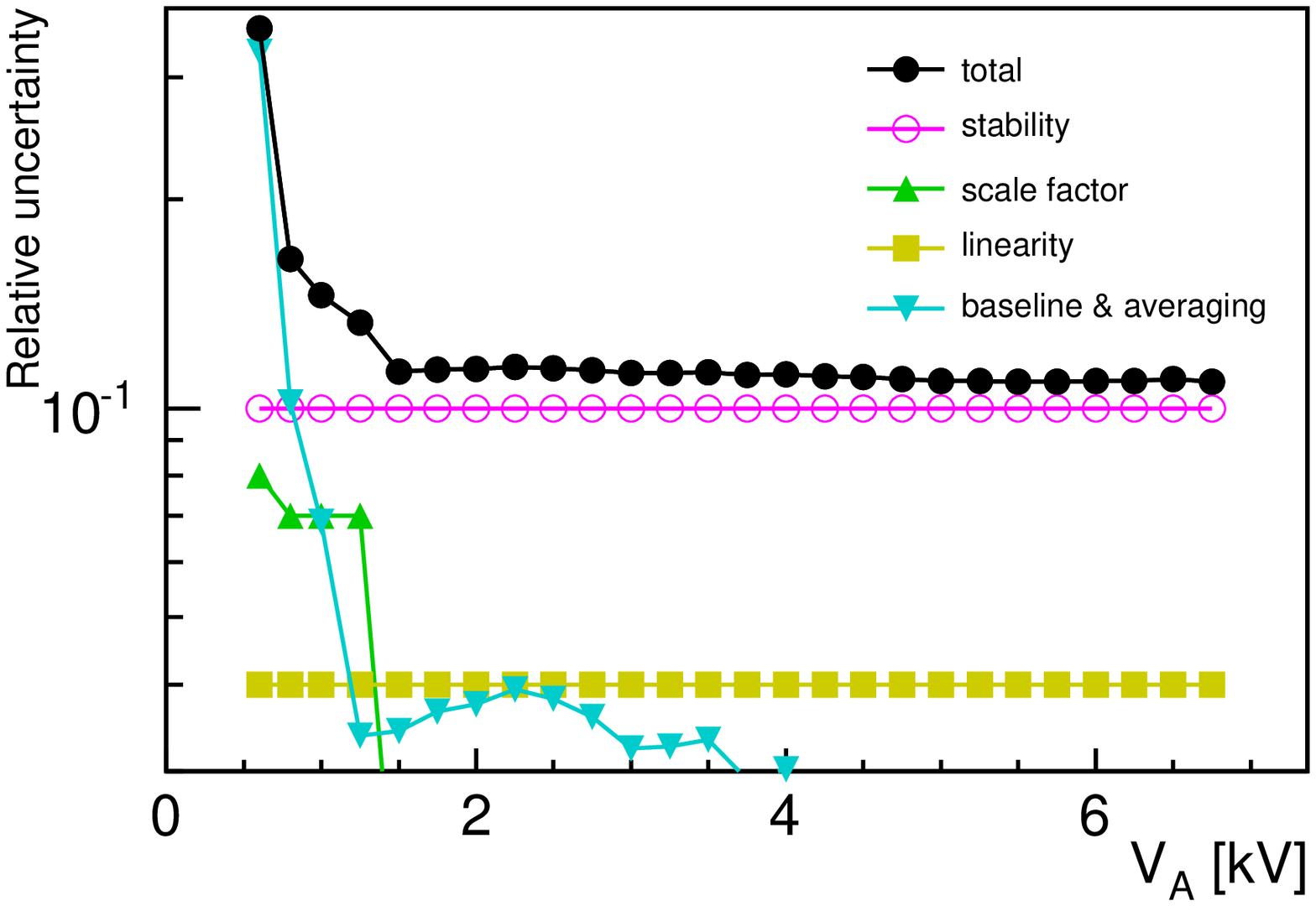} 
  \caption{Systematic uncertainties on the 10~$\mu$m anode wire data.
    Left: uncertainties on the bottom PMT signal as a function of $V_{A}$. 
    Right: uncertainties on the preamplifier signal as a function of $V_{A}$.
    Both uncertainties are dominated by the stability.}
  \label{fig:syst}
 \end{center}
\end{figure}

Figure~\ref{fig:syst} shows the relative uncertainties as a function of $V_{A}$. 
Total relative uncertainties of this type for the 10~$\mu$m anode wire are approximately 15~\% (10~\%) for the PMT (preamplifier) signals.
Both uncertainties are dominated by the stability measurement.
The same uncertainties evaluated for the 5~$\mu$m anode wire diameter are 6~\% and 20~\%, respectively.

A complimentary systematic uncertainty on the fit result is assessed by considering alternate data samples taken slightly under different conditions.
The goal here is to understand how robust the final analysis result is using slightly different experimental conditions.
The fit used to extract the thresholds for charge multiplication and proportional scintillation is re-run using these additional data samples, and the resulting difference is taken as a systematic uncertainty.
The details of this calculation are presented in the next section.

The other major class of systematic uncertainties affecting the analysis is related to the factor ($g$) used to convert the S2 size in photo-electrons to the S2 gain in photons per drift electron ($ph/e^-$).
This conversion factor is defined as
\begin{eqnarray}
g = \frac{1}{\frac{E_{\alpha}}{W} f_{ion} \epsilon_{LC} \epsilon_{Q} f_{LXe} \epsilon_{dy}},
\end{eqnarray}
where $E_{\alpha}$ is the energy of $\alpha$-particles emitted by the $^{210}$Po source, $W$ is the ionization energy of xenon, $f_{ion}$ is the fraction of electrons which escape recombination, $\epsilon_{LC}$ is the bottom PMT S2 LCE, $\epsilon_{Q}$ is the QE of the PMTs at ambient temperature, $f_{LXe}$ is a correction factor for the QE to account for the difference between ambient and LXe temperature, and $\epsilon_{dy}$ is the photo-electron collection efficiency at the first dynode.
The uncertainty on $f_{ion}$ comes from the electric field variation around the needle source, which can change from 1.5 to 3.0~kV/cm.
This uncertainty is estimated by electric field simulation and is dominated by the leakage potential from the anode wire through the gate wires and the geometric effect of the needle's round shape.
The uncertainty on $\epsilon_{LC}$ is obtained by changing various properties in Geant4's optical models.
Specifically, the reflectivities of the PTFE, stainless steel, and gold are varied within uncertainties found in the literature \cite{bib:Yamashita}.\footnote{Reflectivities except for PTFE from Refractive index database: \href{http://refractiveindex.info/}{http://refractiveindex.info/}}
The surface conditions of the PTFE and stainless steel are also varied within parameters defined by Geant4.
The scattering and absorption lengths are similarly changed according to values from the literature \cite{bib:meg}.
A final, small uncertainty is obtained by varying the position of the S2 photon creation.
The uncertainty on the other parameters in $g$ are taken from the literature.
Each factor, and its associated uncertainty, used in the calculation of $g$ is shown in table~\ref{tab:syst}.

\begin{table}[htbp]
  \begin{center}
    \caption{Factors and related uncertainties used to convert the observed S2 signal~(in PE) to the S2 gain~(in $ph/e^-$).} 
\vspace*{8mm}
    \begin{tabular}{cc} \hline\hline
        $E_{\alpha}$ [eV] & $5.41 \times 10^6$ \\
        $W$ [eV] & $15.6 \pm 0.3$  \cite{bib:Takahashi}\\
        $f_{ion}$ & $(4.15 \pm 0.65)\times 10^{-2}$ \cite{Aprile:1991} \\
        $\epsilon_{LC}$  & $0.212^{+0.048}_{-0.015}$ \\
        $\epsilon_{Q}$ & $0.323 \pm 0.040$ \cite{Aprile:2012dy} \\
        $f_{LXe}$ & $1.07 \pm 0.02$ \cite{Aprile:2012dy} \\
        $\epsilon_{dy}$ & $0.750 \pm 0.025$ \\
        \hline
        conversion factor [($ph/e^-)/$PE] & $(1.26^{+0.38}_{-0.27}) \times 10^{-3}$  \\
        \hline\hline
    \end{tabular}
    \label{tab:syst}
  \end{center}
\end{table}


\subsection{Threshold extraction and other results}
\begin{figure}[htbp]
 \begin{center}
  \includegraphics[width=.45\textwidth]{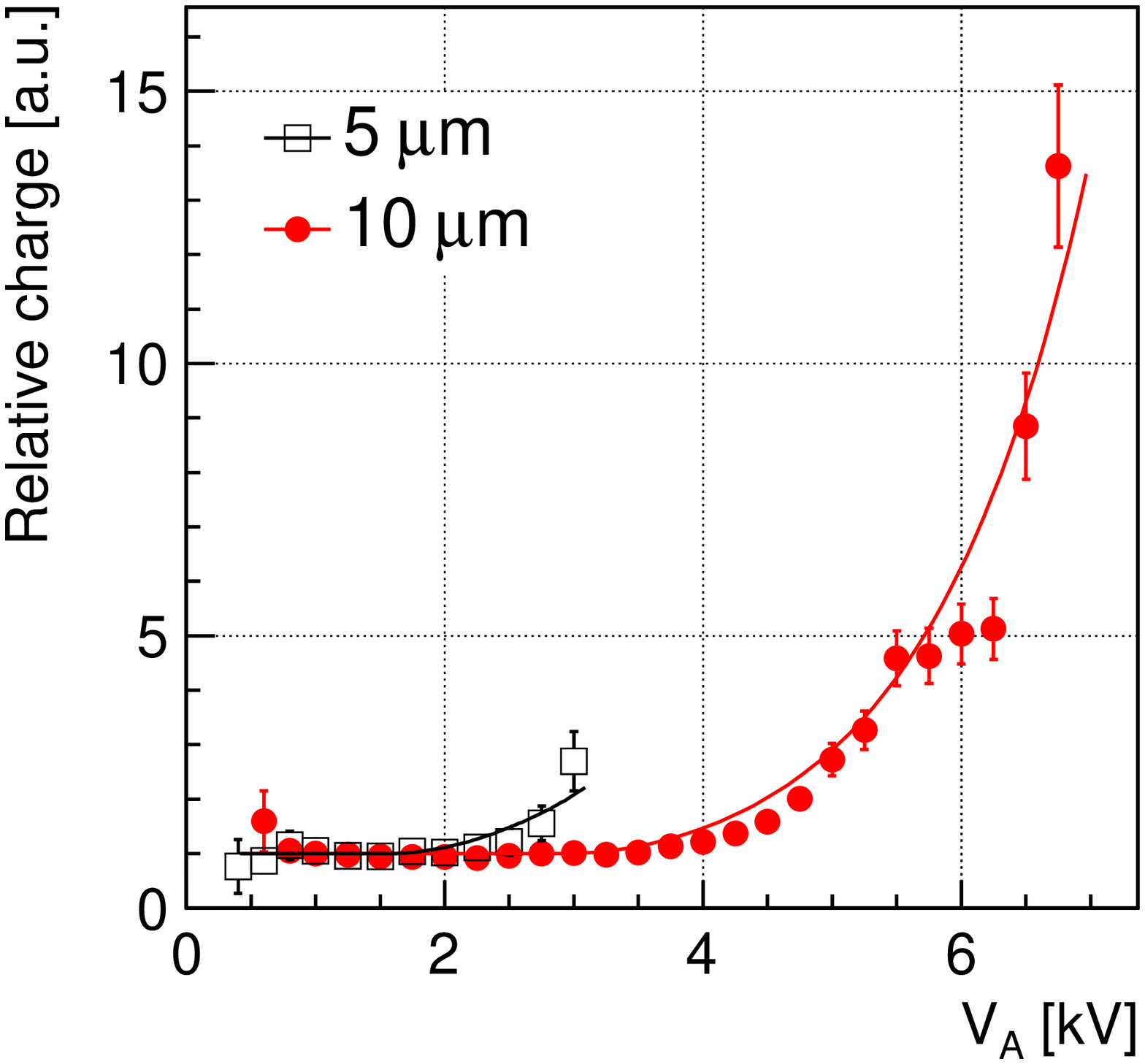} 
  \includegraphics[width=.45\textwidth]{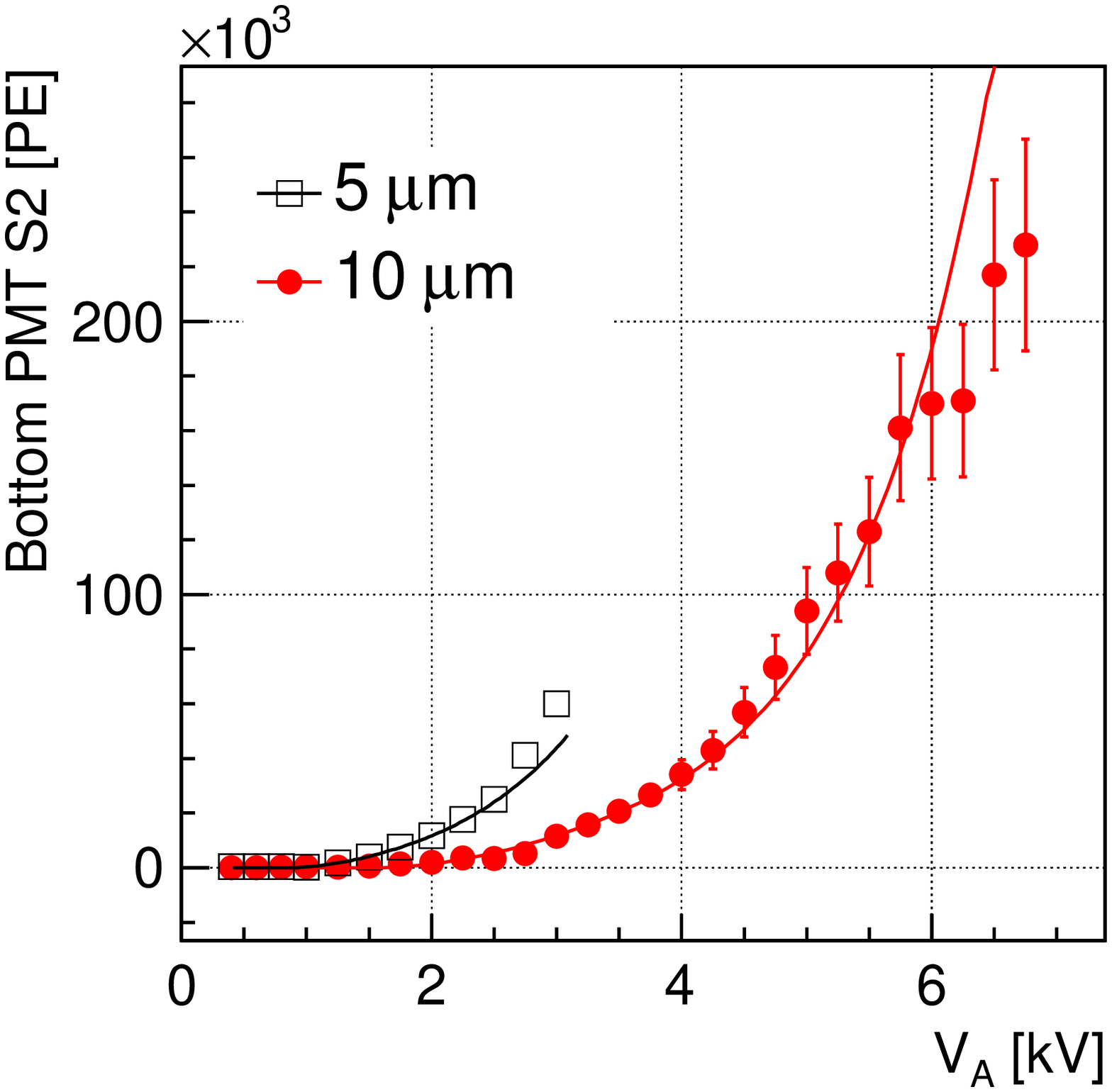} 
  \caption{Left: The preamplifier relative charge signal as a function of $V_{A}$. 
    While the amount of charge is flat at lower voltages, it rapidly increases at higher voltages, indicating the onset of charge multiplication.
    Right: The bottom PMT S2 as a function of $V_{A}$. 
    The lines are from the simultaneous fit of the 10~$\mu$m and 5~$\mu$m data points.}
  \label{fig:main}
 \end{center}
\end{figure}

The primary result from the acquired data is the evolution of the proportional scintillation signal in LXe as a function of the voltage on the anode wire.
Figure~\ref{fig:main} shows the charge signal from the preamplifier~(left), measured relative to the baseline at low $V_{A}$, and the S2 signal from the bottom PMT~(right) measured with the 10 and 5~$\mu$m anode wires.
A maximum average S2 signal of $(2.28 \pm 0.34)\times 10^5$~PEs was obtained with the 10~$\mu$m anode wire at $V_{A}=6.75$~kV.
Using the conversion factor $g$ defined in the previous section, this corresponds to an S2 gain of $287\substack{+97\\-75}$~$ph/e^{-}$, with a factor of $\sim14$ electron multiplication.

To extract the physical properties of electron multiplication and proportional scintillation in LXe, expressions derived for xenon in the gas phase in~\cite{bib:Leo,bib:Bolozdynya} were modified to remove the pressure dependence and were used to fit the charge and S2 signals in LXe:
\begin{eqnarray}
\Delta N_e &=&  N_e \theta_0 \exp \left( -\frac{\theta_1}{E(\vec{x},V_{A},d_w) - \theta_2}  \right) \Delta \vec{x},\\
\Delta N_{\gamma}  &=& N_e \theta_3(E(\vec{x},V_{A},d_w) - \theta_4) \Delta \vec{x},
\end{eqnarray}
where $\theta_i$ are the fit parameters, $E$ is the electric field strength, which is itself a function of the position of the electron ($\vec{x}$), the anode voltage difference ($V_{A}$), and the diameter of the anode wire ($d_w$), and $\Delta \vec{x}$ is the drift path of the electron.
The COMSOL simulation package is used to evaluate $E(\vec{x},V_{A},d_w)$ at each point along the electron's drift path, and numerical integration of equations 4.2 and 4.3 is used to propagate changes in the number of electrons and photons produced during the proportional scintillation process.
$N_e$ is the number of electrons at a given point where the numerical integration is performed.
The number of electrons and photons generated as the electron travels between two points (defined by $\Delta \vec{x}$) is given by $\Delta N_e$ and $\Delta N_{\gamma}$, respectively.
$\Delta N_e$ is then used to update the number of electrons for evaluation at the next step along the drift path.
$\Delta N_{\gamma}$, integrated over the whole electron drift path, gives the total S2 response.
Table~\ref{tab:fit} shows the result of the simultaneous fit to the 10 and 5~$\mu$m data samples, as well as individual fit results for the 10 and 5~$\mu$m wire samples.
The main parameters of interest are $\theta_2$ (the threshold for charge multiplication) and $\theta_4$ (the threshold for S2 production).
$\theta_3$ is also of interest, as it represents the local S2 gain.
The somewhat arbitrary choice of units for the $\theta_3$ is designed to emphasize the characteristic electric field (in kV/cm) and the characteristic length scale (in $\mu$m) for the proportional scintillation process defined by the current experimental setup.

\begin{figure}[htbp]
 \begin{center}
  \includegraphics[width=.45\textwidth]{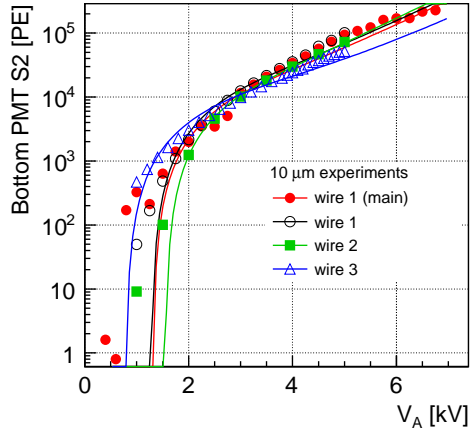}
  \caption{Results using alternate 10$\mu$m data samples, as a cross-check for the baseline fit result for the S2 threshold.
  Two data sets are available with different wire configurations, and one with the same wire taken before the main data taking period began.
  In both cases, the preamplifier was not used, so only information from the PMTs is available.
  The difference in extracted thresholds across these different data samples is used to assess a systematic uncertainty on the S2 production threshold.}
  \label{fig:10um_cross_checks}
 \end{center}
\end{figure}

Positive correlations between the gain and threshold were obtained in the fitting since the threshold at higher electric field requires a higher gain.
While the fits describe the overall features of electron multiplication and proportional scintillation, some discrepancy does remain between the data and this simple model.
In particular, while the threshold of proportional scintillation is consistent for all the fits, other parameters are not consistent.
In addition to the data set presented in this analysis, three other data sets are available using a similar set up.
Two data samples are available with a different 10~$\mu$m wire, and a third with the same wire taken before the main data taking run started.
The preamplifier was not connected to the apparatus for these runs, but it is nonetheless possible to use information from the PMTs taken during these runs.
To assess the systematic uncertainty due to the stability of the detector, these datasets are used to to fit the low $V_{A}$ region ($V_{A} < 5.0$), assuming charge information from the baseline result.
The results of these fits are shown in figure \ref{fig:10um_cross_checks}.

Considering the disagreement of the results from different fits as uncertainties, the thresholds of electron multiplication and proportional scintillation are inferred to be $725\substack{+48\\-139}$ and $412 \substack{+10\\-133}$~kV/cm, respectively. 
The central value was obtained from the 10 and 5~$\mu$m simultaneous fit.
The thresholds correspond to the S2 proportional signal starting at $\sim10$~$\mu$m from the anode wire surface and electron multiplication starting at a few~$\mu$m from the anode wire surface.

\begin{table}[htbp]
  \begin{center}
    \caption{Results from a simultaneous fit of the 10 and 5~$\mu$m data and individual fits of the 10 and 5~$\mu$m data. 
      The last line shows the S2 gain factor after converting PE to $ph/e^-$ using the conversion factor $g$.} 
\vspace*{8mm}
{\small
    \begin{tabular}{lrrrr} \hline\hline

      parameter                                           & 10 \& 5~$\mu$m  & only 10~$\mu$m  & only 5~$\mu$m   \\
        \hline
        $\theta_0$: charge gain factor [1/($\mu$m$\cdot e^-$)] & $0.80 \pm 0.10$ & $1.15 \pm 0.15$ & $1.46 \pm 0.02$ \\
        $\theta_1$: slope in charge gain [kV/cm]               & $242 \pm 45$    & $561 \pm 119$    & $298 \pm 1$     \\
        $\theta_2$: threshold of charge mult. [kV/cm]          & $725 \pm 48$    & $586 \pm 47$    & $750 \pm 1$     \\
        $\theta_3$: S2 gain factor [PE/(kV/cm$\cdot \mu$m)]    & $16.6 \pm 1.1$  & $13.3 \pm 0.4$  & $17.9 \pm 3.4$  \\
        $\theta_4$: threshold of S2 [kV/cm]                    & $412 \pm 10$    & $399 \pm 7$     & $416 \pm 13$    \\
        $\chi^2$/ndf                                      & $125/63$        & $71.4/42$       & $19.9/16$       \\
        \hline
        $\theta_3$: S2 gain factor [$ph/e^-$/(kV/cm$\cdot \mu$m)]  & $(2.09^{+0.65}_{-0.47})\times10^{-2}$ & $(1.68^{+0.51}_{-0.36})\times10^{-2}$ & $(2.26^{+0.80}_{-0.65})\times10^{-2}$ \\
        \hline\hline
    \end{tabular}
}
    \label{tab:fit}
  \end{center}
\end{table}

\begin{figure}[htbp]
 \begin{center}
  \includegraphics[width=.4\textwidth]{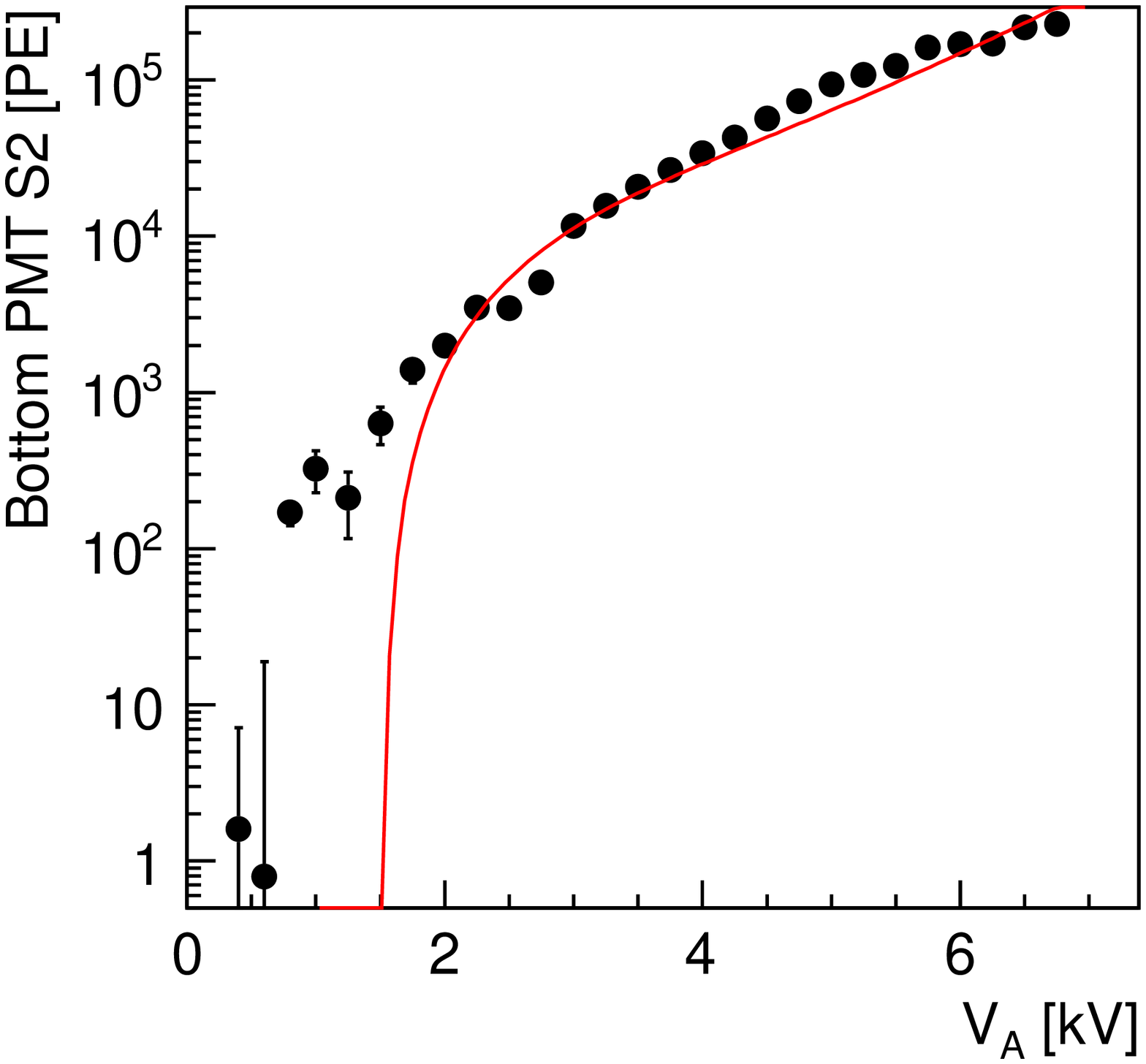} 
  \includegraphics[width=.45\textwidth]{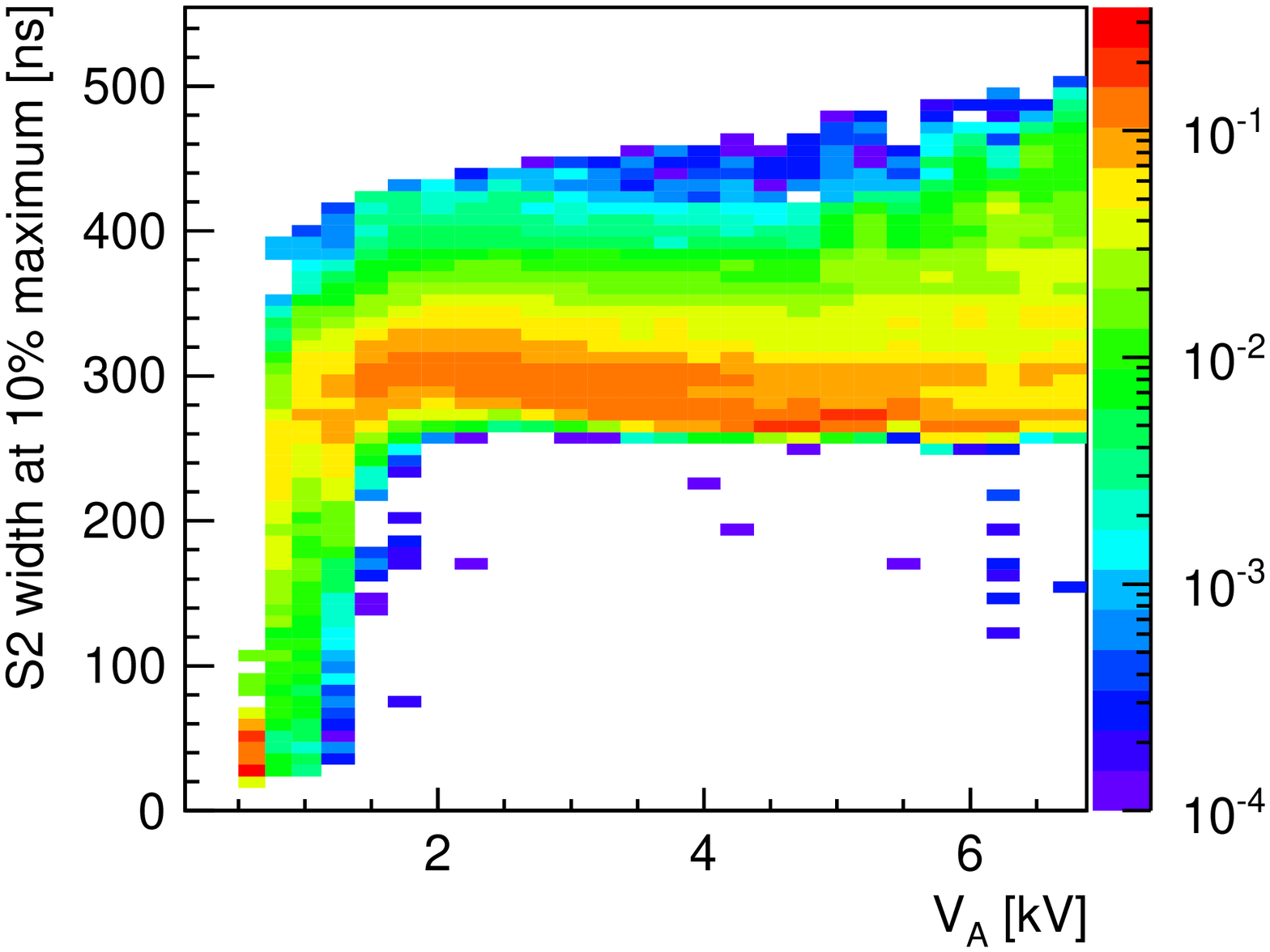} 
  \caption{Left: S2 as a function of $V_{A}$ from the 10~$\mu$m anode wire.
    Significant disagreement between the data and the fit can be seen at low anode voltages.
    Right: S2 pulse width at 10~\% of pulse height. 
    The colors represent number of events normalized at each $V_{A}$. 
    The S2 width decreases at low voltages.}
  \label{fig:width}
 \end{center}
\end{figure}

Proportional scintillation was observed at $V_{A}$ even below the S2 threshold.
The fit disagrees significantly with the data at lower $V_{A}$ as shown in figure~\ref{fig:width}~(left).
The S2 width at 10~\% pulse height drops at these voltages as shown in figure~\ref{fig:width}~(right).
An S2 width of 300~ns is consistent with the expectation that the electron cloud is spread by longitudinal and transverse diffusion~\cite{Aprile:2010}.
One hypothesis for the S2 width reduction at lower $V_{A}$ is that only some fraction of electrons acquire energy by the electric field and emit S2 light, and therefore the actual threshold is not determined by the electric field.

\begin{figure}[htbp]
 \begin{center}
  \includegraphics[width=.45\textwidth]{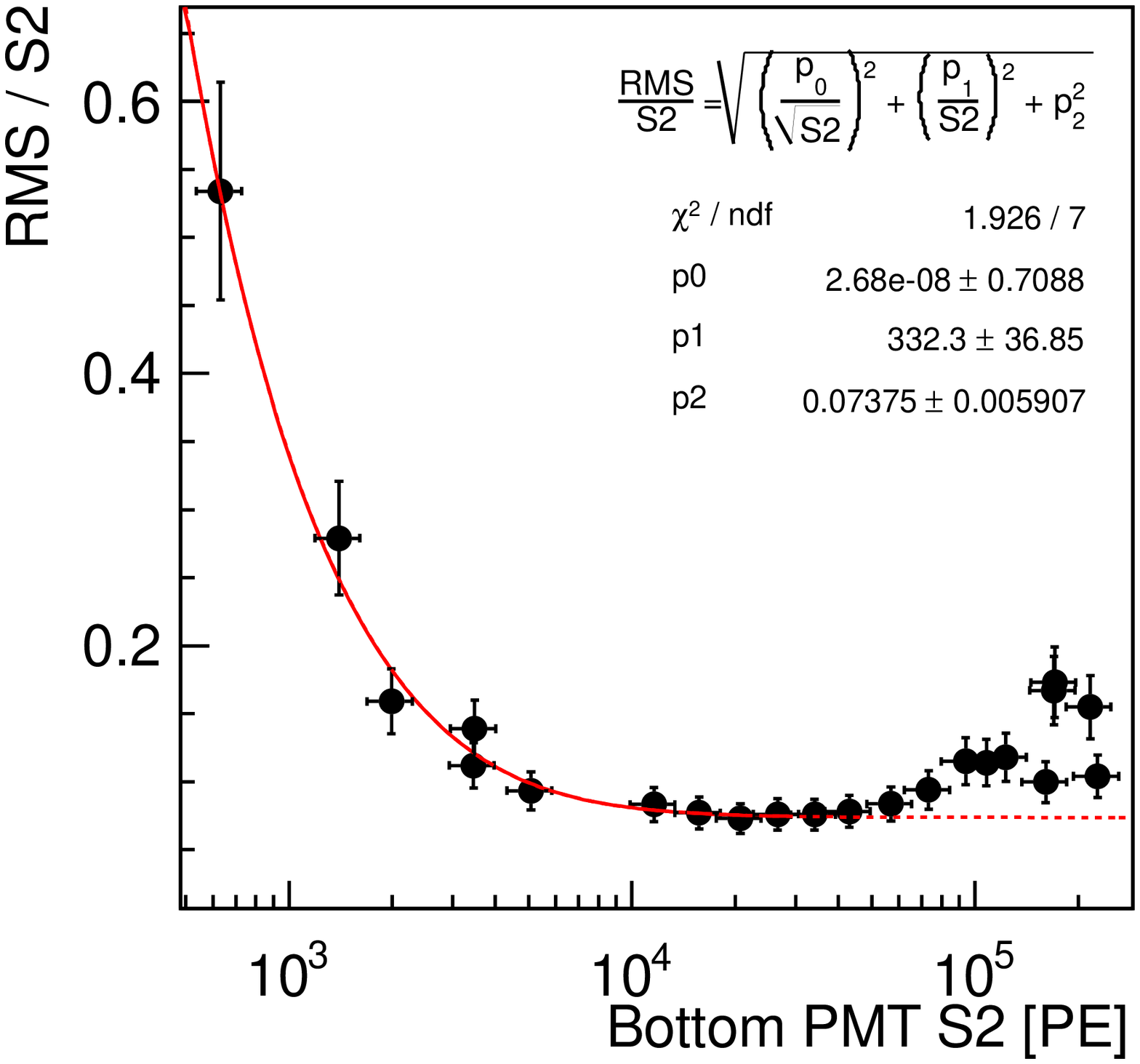}
  \includegraphics[width=.45\textwidth]{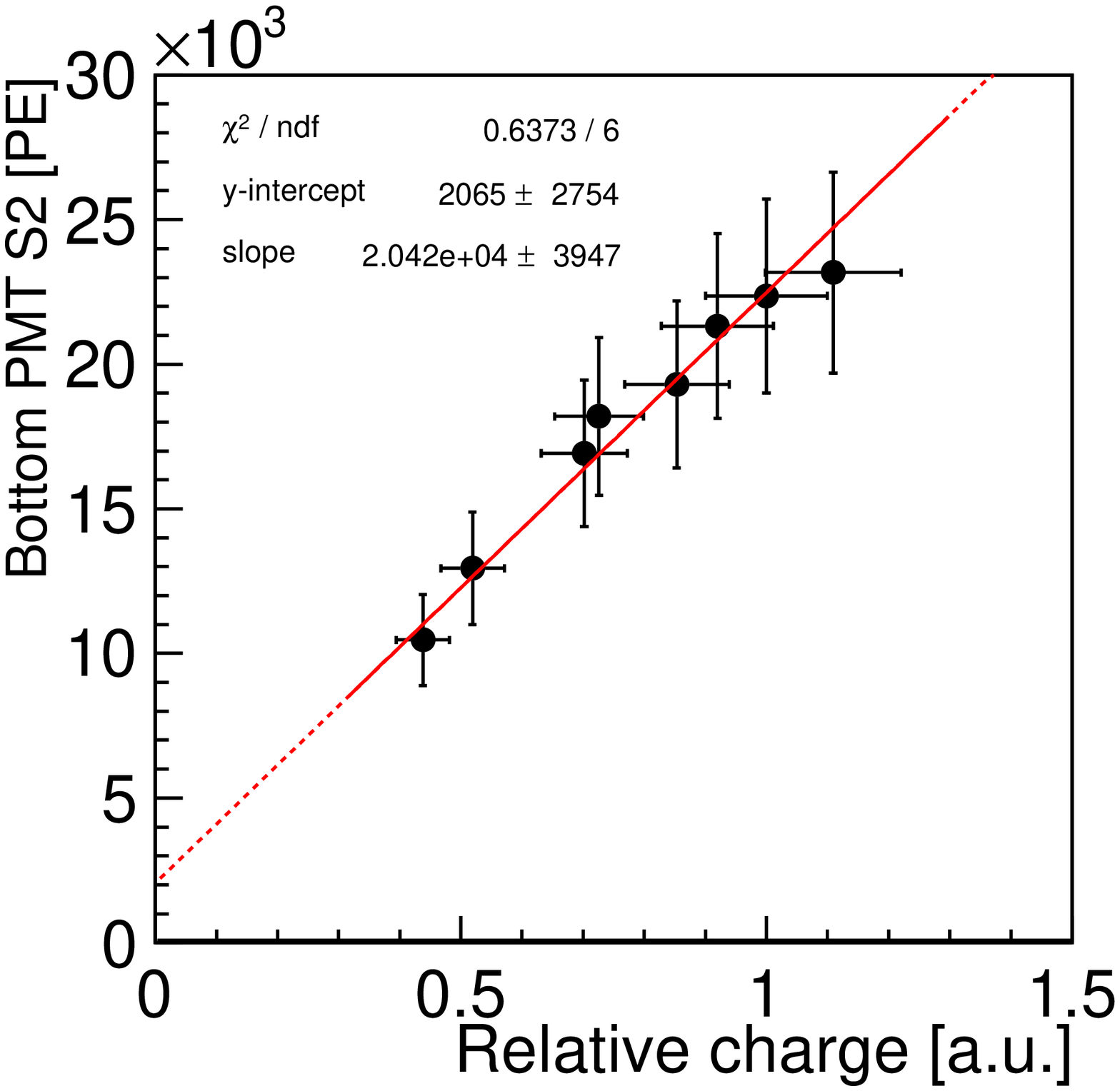}
  \caption{Left: S2 resolution as a function of S2 using 10~$\mu$m anode wire. 
    Right: proportionality test of the S2 to the number of electrons observed by the preamplifier, using the 10~$\mu$m anode wire.}
  \label{fig:misc}
 \end{center}
\end{figure}

The S2 resolution as a function of $V_{A}$ from the standard event selection is shown in figure~\ref{fig:misc}~(left).
The best S2 resolution is 7~\% RMS for an S2 of $\sim 2 \times 10^4$~PE, which corresponds to $V_{A}=3.5$~kV.
The fractional S2 resolution is fit with a formula containing terms which correspond to the various sources of uncertainty which limit the final resolution.
The constant term of this fit indicates that the first type of systematic uncertainties is overestimated.
The degradation of the resolution at the higher $V_{A}$ is ascribed to the onset of electron multiplication.

To address the proportionality of the S2 signal in LXe with the number of drift electrons, a measurement was performed at fixed $V_{A}=3.5$~kV for different values of $V_{C}$.
Changes in $V_{C}$ only affect $f_{ion}$, so an indication of proportional scintillation would be the linearity of the size of the observed S2 with the number of electrons measured by the preamplifier.
This linearity is shown in figure~\ref{fig:misc}~(right).

\section{Discussion}
\label{sec:dis}

In the work presented here, we have studied the properties of proportional scintillation in LXe using thin wires.
This work represents an important confirmation of previous studies, as well as a step forward in our understanding of this phenomenon.
Using a simple theoretical model for the onset of proportional scintillation and charge multiplication, we extract the thresholds for each process using a fit to data.

The threshold of $412\substack{+10\\-133}$~kV/cm for proportional scintillation in LXe estimated from this study is consistent with the result obtained in \cite{bib:Masuda1}.
If LXe is considered as a gas at a pressure of 520 atm, as discussed in \cite{bib:Masuda1}, it is possible to compare the S2 gain and proportional scintillation threshold as measured from the data to theoretical values presented in \cite{bib:Bolozdynya}, which are derived only for the gas phase.
Under this assumption, the theoretical calculation in \cite{bib:Bolozdynya} implies a gain of 70~$ph/e^-/$(kV/cm$\cdot$cm) and a threshold of 520~kV/cm.
While the measured threshold of $412\pm10$~kV/cm agrees with this theoretical value, the S2 gain of $209\substack{+65\\-47}$~$ph/e^-/$(kV/cm$\cdot$cm) is more than a factor of three higher than the expected value, implying that this theoretical model is useful but not completely predictive.

Assuming the validity of these results based on agreement with other measurements and theoretical calculations, it is instructive to ask about their implication for future detectors.
One question one can ask is whether there exists an optimal wire diameter for the production of proportional scintillation in LXe.
The  observation of continuous photon emission limited $V_{A}$ to a value corresponding to an electric field strength $\geq1.6$~MV/cm on the anode wire surface.
Thus, to avoid continuous light emission, one needs a lower electric field strength.
One other practical consideration is the total available high voltage supply.
Our power supply enabled a maximum voltage ($V_{A}$) of 8 kV to be applied.
Assuming an electric field strength on the wire surface of 1.5~MV/cm and all other experimental conditions as in our setup, the results presented in this work imply a maximum S2 gain of $\sim$720~$ph/e^-$, which could be realized using a 15~$\mu$m wire.
If higher voltages are accessible, a larger diameter could be used to realize an even higher S2 gain.
More generally, future experiments can use the results presented in this work to estimate the optimal wire size for their particular setup.

Another important question connected to the feasibility of using thin wires in a single-phase TPC is how sensitive the TPC is for signals composed of a few electrons.
S2 signals comprised of less than 10 electrons are important for the study of WIMPs with a mass down to a few hundred MeV.
The observed maximum S2 gain of $287\substack{+97\\-75}$ photons per electron from this study indicates that it is possible to detect a signal from a few drift electrons.
This gain is similar to that obtained in the gas phase by a dual-phase TPC such as that of the XENON100 experiment, which demonstrated a gain of 454 photons per electron \cite{bib:Aprile_se}.
Even without further optimization, this implies the possibility to utilize proportional scintillation with thin wires in LXe for next generation dark matter detectors, though it is important to note that other applications are possible.

While our work shows the potential of a single-phase TPC for rare events search, more investigations are needed to find the optimal solution to the challenges presented by the use of delicate thin wires. 
A single phase TPC allows for a better S1 light collection efficiency but worse S2 resolution, due to electron multiplication which will occur at the operating conditions required for a sizable S2 signal.
Additionally, the sensitive region of a single-phase TPC may be subdivided using multiple electrodes, thus reducing the high voltage requirement.
The results presented in this work are a first attempt to evaluate the practical application of proportional scintillation in LXe for a next generation, large volume TPC for dark matter direct detection as well as for applications in other areas of research, but more detailed studies are needed to assess the relative merit of these various considerations before deployment of a single-phase TPC is to be undertaken.

\acknowledgments
The authors wish to thank Dr. Karl L. Giboni for the initial idea to use proportional scintillation in LXe for dark matter direct detection experiments, and for his encouragement of this study. 
The authors also gratefully acknowledge Rosie Hood, a visiting student from Imperial College, UK, for her work on the experiment during the summer of 2013.  
This work was carried out with the support of the National Science Foundation for the XENON Dark Matter project at Columbia and Rice Universities.


\newpage
\bibliography{ref}

\begin{thebibliography}{9}
\bibitem{Aprile:2012nq} E.~Aprile et al. (XENON100 Collaboration), \emph{Dark Matter Results from 225 Live Days of XENON100 Data}, \href{http://dx.doi.org/10.1103/PhysRevLett.109.181301}{\emph{Phys.\ Rev.\ Lett.}  {\bf 109} (2012) 181301}.
\bibitem{Akerib:2013tjd} D.~S.~Akerib et al. (LUX Collaboration), \emph{First results from the LUX dark matter experiment at the Sanford Underground Research Facility}, \href{http://dx.doi.org/10.1103/PhysRevLett.112.091303}{\emph{Phys.\ Rev.\ Lett.} {\bf 112} (2014) 091303}.
\bibitem{Aprile:2010} E.~Aprile and T.~Doke, \emph{Liquid xenon detectors for particle physics and astrophysics}, \href{http://dx.doi.org/10.1103/RevModPhys.82.2053}{\emph{Rev.\ Mod.\ Phys.} {\bf 82} (2010) 2053}. 
\bibitem{Chepel:2013} V.~Chepel and H.~Ara\'ujo, \emph{Liquid noble gas detectors for low energy particle physics}, \href{http://dx.doi.org/10.1088/1748-0221/8/04/R04001}{ \emph{JINST} {\bf 8} (2013) R04001}. 
\bibitem{Giboni:2011} Private comminication with K.~Giboni. The idea can be found at {\href{http://rd.kek.jp/slides/20111124/Karl_Giboni.pdf}{KEK seminar}}.
\bibitem{bib:Lansiart1} A.~Lansiart et al., \emph{DEVELOPMENT RESEARCH ON A HIGHLY LUMINOUS CONDENSED XENON SCINTILLATOR}, \href{http://dx.doi.org/10.1016/0029-554X(76)90824-7}{ \emph{Nucl.\ Instrum.\ Methods} {\bf 135} (1976) 47}.
\bibitem{bib:Masuda1} K.~Masuda et al., \emph{A liquid xenon proportional scintillation counter}, \href{http://dx.doi.org/10.1016/0029-554X(79)90600-1}{ \emph{Nucl.\ Instrum.\ Methods} {\bf 160} (1979) 247}.
\bibitem{bib:Derenzo1} S.E.~Derenzo et al., \emph{Electron avalanche in liquid xenon}, \href{http://dx.doi.org/10.1103/PhysRevA.9.2582}{ \emph{Phys.\ Rev.\ A} {\bf 9} (1974) 2582}.
\bibitem{bib:Policarpo1} A.P.L.~Policarpo et al., \emph{Observation of electron multiplication in liquid xenon with a microstrip plate}, \href{http://dx.doi.org/10.1016/0168-9002(95)00457-2}{ \emph{Nucl.\ Instrum.\ Methods} {\bf A365} (1995) 568}.
\bibitem{bib:Arazi} L.~Arazi et al., \emph{First observation of liquid-xenon proportional electroluminescence in THGEM holes}, \href{http://dx.doi.org/10.1088/1748-0221/8/12/C12004}{\emph{JINST} {\bf 8} (2013) C12004}. 
\bibitem{bib:Buzulutskov} A.~Buzulutskov, \emph{Advances in Cryogenic Avalanche Detectors}, \href{http://dx.doi.org/10.1088/1748-0221/7/02/C02025}{\emph{JINST} {\bf 7} (2012) C02025}. 
\bibitem{hamamatsu} Model R8520-406 SEL from {\href{http://www.hamamatsu.com/jp/en/index.html}{Hamamatsu Photonics K.K.}}
\bibitem{Aprile:2011dd} E.~Aprile et al., \emph{The XENON100 Dark Matter Experiment}, {\emph{Astropart.Phys.} {\bf 35} (2012) 573-590}.
\bibitem{Aprile:2012dy} E.~Aprile et al., \emph{Measurement of the Quantum Efficiency of Hamamatsu R8520 Photomultipliers at Liquid Xenon Temperature}, \href{http://dx.doi.org/10.1088/1748-0221/7/10/P10005}{ \emph{JINST} {\bf 7} (2012) 1005}.
\bibitem{CAENHV}{Model SY4527 from \href{http://www.caen.it/}{CAEN S.p.A.}}
\bibitem{AMPTEK}{Model A225 from \href{http://www.amptek.com}{Amptek, Inc.}}
\bibitem{bib:Guillaume1} G.~Plante et al., \emph{New measurement of the scintillation efficiency of low-energy nuclear recoils in liquid xenon}, \href{http://link.aps.org/doi/10.1103/PhysRevC.84.045805}{\emph{Phys.\ Rev.\ C} {\bf 84} (2011) 045805}.
\bibitem{bib:Kyungun1} E.~Aprile et al., \emph{Measurement of the Scintillation Yield of Low-Energy Electrons in Liquid Xenon}, \href{http://dx.doi.org/10.1103/PhysRevD.86.112004}{\emph{Phys.\ Rev.\ D} {\bf 86} (2012) 112004}.
\bibitem{LAKESHORE}{Model 340 from \href{http://www.lakeshore.com/}{Lake Shore Cryotronics, Inc.}}
\bibitem{SAES} {Model PS3-MT3-R-1 from \href{http://www.saespuregas.com/Home.html}{SAES Pure Gas, Inc.}}
\bibitem{bib:Takahashi} T.~Takahashi et al., \emph{Average energy expended per ion pair in liquid xenon}, \href{http://dx.doi.org/10.1103/PhysRevA.12.1771}{\emph{Phys.\ Rev.} {\bf A12} (1975) 1771}.
\bibitem{Aprile:1991} E.~Aprile et al., \emph{Ionization of liquid xenon by $^{241}$Am and $^{210}$Po alpha particles}, \href{http://dx.doi.org/10.1016/0168-9002(91)90138-G}{\emph{Nucl.\ Instrum.\ Methods} {\bf A307} (1991) 119}.
\bibitem{bib:Miller} L.S.~Miller, S.~Howe, and W.E.~Spear, {\it et al.}, \emph{Charge Transport in Solid and Liquid Ar, Kr, and Xe} \href{http://dx.doi.org/10.1103/PhysRev.166.871}{\emph{Phys.\ Rev.}, {\bf 166} (1968) 871}.
\bibitem{CAEN}{Model v1724 from \href{http://www.caen.it/}{CAEN S.p.A.}}
\bibitem{LECROY}{Model 623B from \href{http://teledynelecroy.com/}{Teledyne LeCroy}}
\bibitem{BNC}{Model PB-4 from \href{http://www.berkeleynucleonics.com/}{Berkeley Nucleonics Corporation}}
\bibitem{SRS}{Model DS345 from \href{http://www.thinksrs.com/}{Stanford Research Systems}}
\bibitem{comsol} {\href{http://www.comsol.com/}{COMSOL Inc.}}
\bibitem{geant4} S.~Agostinelli et al., \emph{Geant4 - a simulation toolkit}, \href{http://dx.doi.org/10.1016/S0168-9002(03)01368-8}{ \emph{Nucl.\ Instrum.\ Methods} {\bf A506} (2003) 250}.
\bibitem{bib:Yamashita} M.~Yamashita et al., \emph{Scintillation response of liquid Xe surrounded by PTFE reflectors for gamma rays} \href{http://dx.doi.org/10.1016/j.nima.2004.06.168}{\emph{Nucl.\ Instr.\ Methods\ Phys.\ Res.}, {\bf 535} (2004) 692}.
\bibitem{bib:meg} A.~Baldini et al., \emph{Absorption of scintillation light in a 100 $\ell$ liquid xenon $\gamma$-ray detector and expected detector performance} \href{http://dx.doi.org/10.1016/j.nima.2005.02.029}{\emph{Nucl.\ Instrum.\ Methods} {\bf A545} (2005) 753}.
\bibitem{bib:Leo} W.R.~Leo, \emph{Techniques for Nuclear and Particle Physics Experiments}, Springer-Verlag, New York 1993.
\bibitem{bib:Bolozdynya} A.I.~Bolozdynya, \emph{Two-phase emission detectors and their applications}, \href{http://dx.doi.org/10.1016/S0168-9002(98)00965-6}{\emph{Nucl.\ Instrum.\ Methods} {\bf A422} (1999) 314}. 
\bibitem{bib:Aprile_se} E.~Aprile et al., (XENON100 Collaboration), \emph{Observation and applications of single-electron charge signals in the XENON100 experiment}, {\emph J. Phys. G: Nucl. Part. Phys.} {\bf 41} (2014).
\end{thebibliography}

\end{document}